\documentclass[twocolumn]{aastex7}
\usepackage{graphicx}
\usepackage{gensymb}
\usepackage{longtable}
\usepackage{amsmath}
\def\kms{$\rm km~s^{-1}$}

\begin{document}

%\title{Near Infrared HST Images of Pa 30, the Remnant of the Galactic Supernova of 1181 AD}

\title{Near Infrared HST Images Resolve the Filamentary Emission Knots in Pa 30, \\
the Remnant of the Galactic Supernova of 1181 AD}

%\title{Discovery of an Exceptional Optical Nebulosity in the Suspected Galactic SN Iax Remnant Pa 30 Linked to the Historical Guest Star of 1181 CE}

\author[0000-0003-3829-2056]{Robert A.\ Fesen}
\email{robert.fesen@dartmouth.edu}
\affiliation{6127 Wilder Lab, Department of Physics and Astronomy, Dartmouth College, Hanover, New Hampshire, 03755, USA}

\author[0000-0002-0763-3885]{Dan Milisavljevic}
\email{dmilisav@purdue.edu}
\affiliation{Purdue University, Department of Physics and Astronomy, 525 Northwestern Ave, West Lafayette, IN 47907, USA}
\affiliation{Integrative Data Science Initiative, Purdue University, West Lafayette, IN 47907, USA}

\author[0000-0003-4069-2817]{Peter Garnavich}
\email{pgarnavi@nd.edu}
\affiliation{Department of Physics and Astronomy, University of Notre Dame, Notre Dame, IN 46556, USA}

\author[0000-0002-7507-8115]{Dan Patnaude}
\email{dpatnaude@cfa.harvard.edu}
\affiliation{Center for Astrophysics, 60 Garden Street, Cambridge, MA 02138, USA}

\author[0000-0001-7380-3144]{Tea Temim}
\email{tea.temim@columbia.edu}
\affiliation{Department of Astronomy, Columbia University, 550 West 120th Street, New York, NY 10027-6601, USA}

\begin{abstract}
We report the results of a near-infrared HST WFC3/IR imaging program investigating the sub-arcsecond structure of optical filaments in the Galactic nebula, Pa~30, believed to be a SN~Iax remnant of the historic supernova of 1181 AD.
Using a near infrared filter sensitive to 
[\ion{S}{2}] and [\ion{S}{3}] emission lines,
a 1.2 $\mu$m filter to reject stars, and a
deep Gemini GMOS [\ion{S}{2}] image to help define the remnant's filaments, 
we identify 226 knots along radially aligned filaments in 15 fields.
Knot median PSF-corrected FWHM 
is $0.154''$ ($\simeq$350 AU at 2.3 kpc) 
with 32 (14\%) 
unresolved at a PSF of $0.225''$.
We find that Pa 30’s knots are not round but elongated along
the filaments. Fitting stacks of knots with their axes aligned to
their filament directions and calibrating with artificial knots gives intrinsic
sizes of  $0.18''$ along and $0.11''$ across 
filaments, with about 75\%
of knots longer when measured along the filament.
The minimum nearest-neighbor spacing of knots along a filament is typically
$0.6''$ ($\sim 1400$ AU) so several times their own
size, but can be one arc-second or larger in places. This results in $5''-15''$ long discontinuous chains of separate clumps rather than a continuous structure.
We find excellent agreement with recent ground-based Gemini images of Pa~30
when the HST images are blurred to the 
$\simeq0.5''$ Gemini PSF and conclude that filamentary knots detected in Gemini data
are actually blurred HST resolved knots.
Finally, we give estimates on the electron density and the mass of sulfur in these knots, set an upper limit on their
[\ion{Fe}{2}] 1.257 $\mu$m flux, and briefly discuss implications for modeling the formation of the remnant's filaments.

\end{abstract}
\bigskip

%*******************************

\section{Introduction}

The recently recognized Galactic nebula, Pa~30 (G123.1+4.6) roughly three arc minutes in diameter, is now widely
viewed as the likely supernova remnant (SNR) associated with the
historical guest star reported by Chinese, Japanese and Egyptian astronomers in early August of 1181 CE \citep{Ritter2021,Lykou2023,Schaefer2023,Fisher2025}. The star had an estimated peak brightness $\sim$ 0 mag and  was visible for about six months
\citep{Stephenson1971,Clark1978,Stephenson2002, Schaefer2023} 

Initially identified as a possible planetary nebula, number 30 in the list of possible PN discoveries
by D.\ Patchick \citep{Kronberger2014}, Pa 30 was subsequently found associated with a peculiar infrared star, IRAS00500+6713, which has an unusually high temperature $\sim$210,000 K and a 16,000 km s$^{-1}$ velocity wind \citep{Gvara2019,Garnavich2020}.

%%**********************************
%%%Figure 1: Main Image
%%%%**********************************
\begin{figure*}[ht]
\begin{center}
\includegraphics[angle=0,width=15.0cm]{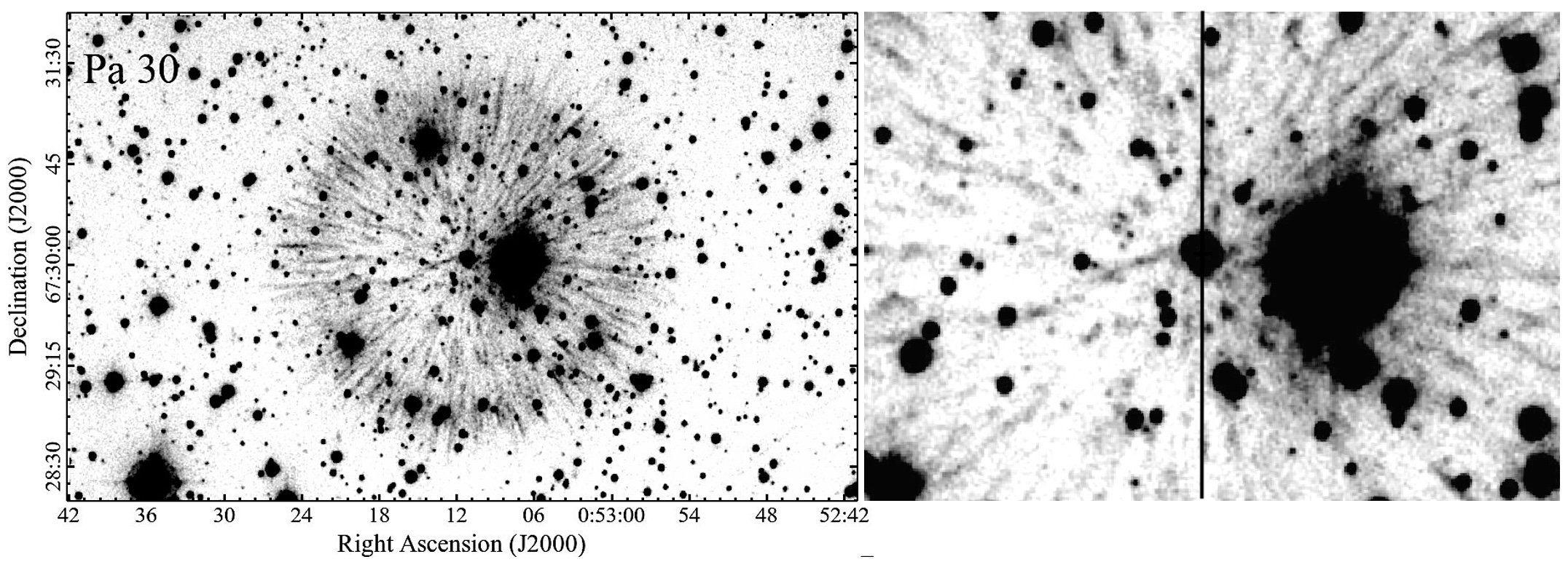} \\
\includegraphics[angle=0,width=15.0cm]{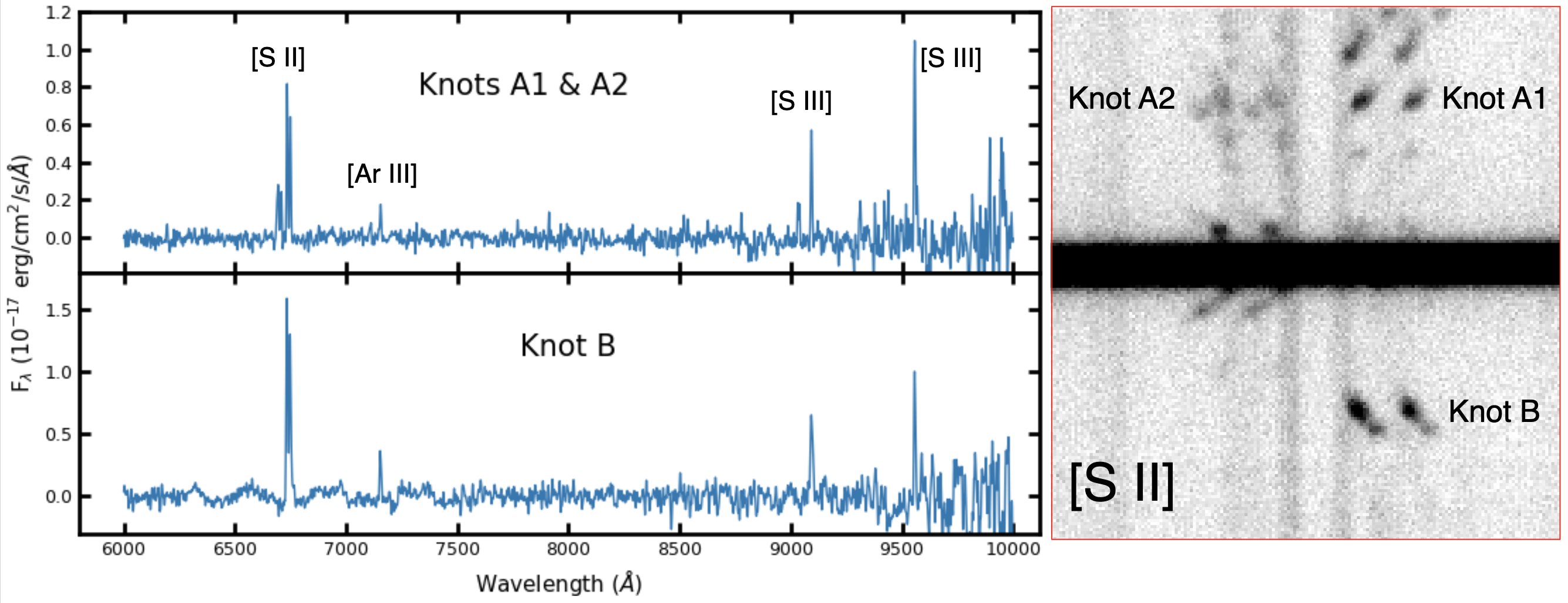}
\caption{Top: A [\ion{S}{2}] $\lambda\lambda$6716, 6731 image of Pa~30 (left panel) with the location where the LBT MODS spectrograph slit was placed (right panel)
Bottom: LBT low-dispersion 1D spectra
for three Pa~30 emission knots (see right panel) showing the presence of 
strong [\ion{S}{3}] $\lambda\lambda$9069, 9531 emissions in addition to the 
[\ion{S}{2}] lines. The horizontal black line is Pa~30's central star.       \label{fig1}
}
\end{center}
\end{figure*}
%%%%%%%*********************************

While no radio emission has been detected from this relatively young Galactic remnant \citep{Shao2025}, the Pa~30 nebula is fairly bright in the mid-infrared 
\citep{Gvara2019,Ritter2021} and has been detected in X-rays \citep{Oskinova2020}.

A relatively young SNR with a hot surviving star raised the possibility that Pa~30 may be the remnant of a double-degenerate merger leading to a rare subtype Ia SN event known as a SN~Iax \citep{Gvara2019,Kashiyama2019,Oskinova2020,Ritter2021,Lykou2023}. Models of WD mergers have suggested that after merger, the star is subject to
magnetohydrodynamic instabilities which can lead to viscous evolution on relatively short time-scales compared to its thermal evolution. The merged star, possibly from ONe + CO 
WDs, eventually leads to either a single, high-mass O/Ne WD, where the majority of the CO WD is ejected \citep{Piro2026}, or will undergo a merger-induced collapse to a neutron star \citep{Shen2012,Schwab2016,Schwab2021}.
Several X-ray, infrared, and optical properties of Pa~30 and its central star suggest it is the product of double WD merger \citep{Yao2023,Ko2024}, although a lack of associated radio emission might challenge its SNR nature \citep{Shao2025}.

Optical spectra of Pa 30 show only emission lines
of S, O, and Ar with no H or He line emissions detected
\citep{Gvara2019,Ritter2021,Lykou2023,Fesen2023}. 
Because of a fairly high estimated interstellar 
extinction ($E(B-V) \simeq 0.88$; \citealt{Ritter2021,Lykou2023}), the remnant's brightest optical line emissions are seen at longer wavelengths, 
with [\ion{O}{3}] $\lambda\lambda$4959,5007 only weakly detected.
Spectra show the [\ion{S}{2}] 
$\lambda\lambda$6716,6731 lines dominate its optical spectra and indicate the 
remnant optical nebula to have an expansion velocities ranging from 
$\approx$600 km s$^{-1}$  up to 
$\approx$1400 km s$^{-1}$
\citep{Ritter2021,Fesen2023,Cunningham2024}.

Although initial optical [\ion{O}{3}] images of Pa~30 showed only a diffuse 
emission structure 
\citep{Kronberger2014,Ritter2021}, subsequent deeper images revealed it to have a remarkably filamentary structure unlike any other Galactic SNR,
with dozens of long, narrow and radially aligned [\ion{S}{2}] 
$\lambda\lambda$6716,6731 emission line filaments with typical lengths between $5''$ and $15''$ 
(see Fig. 1; \citealt{Fesen2023}). 

%%%**********************************
%%%Figure 2: WFC3 observing details
%%%%**********************************
\begin{figure*}[ht]
\begin{center}
\includegraphics[angle=0,width=5.5cm]{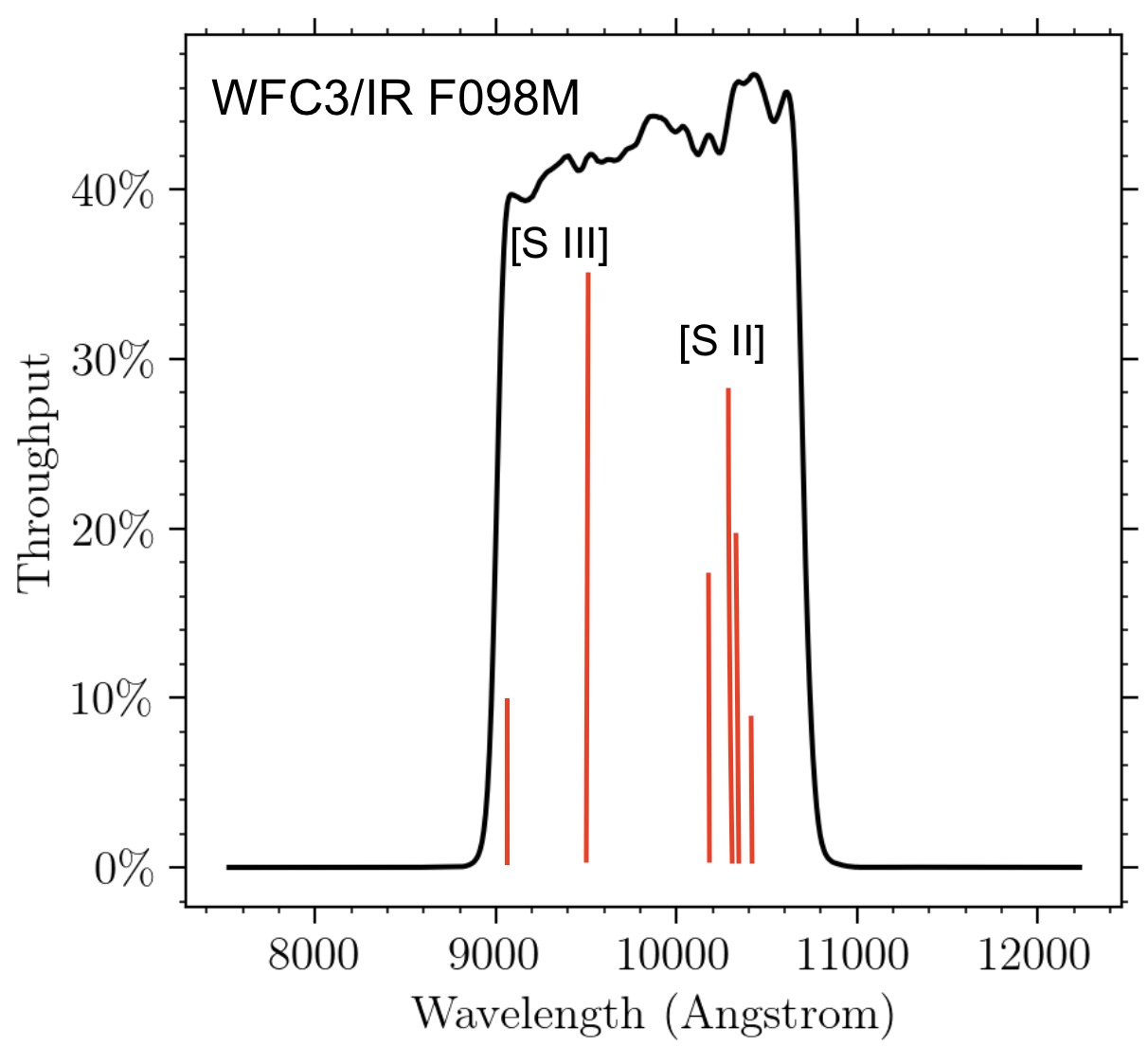} 
\includegraphics[angle=0,width=5.5cm]{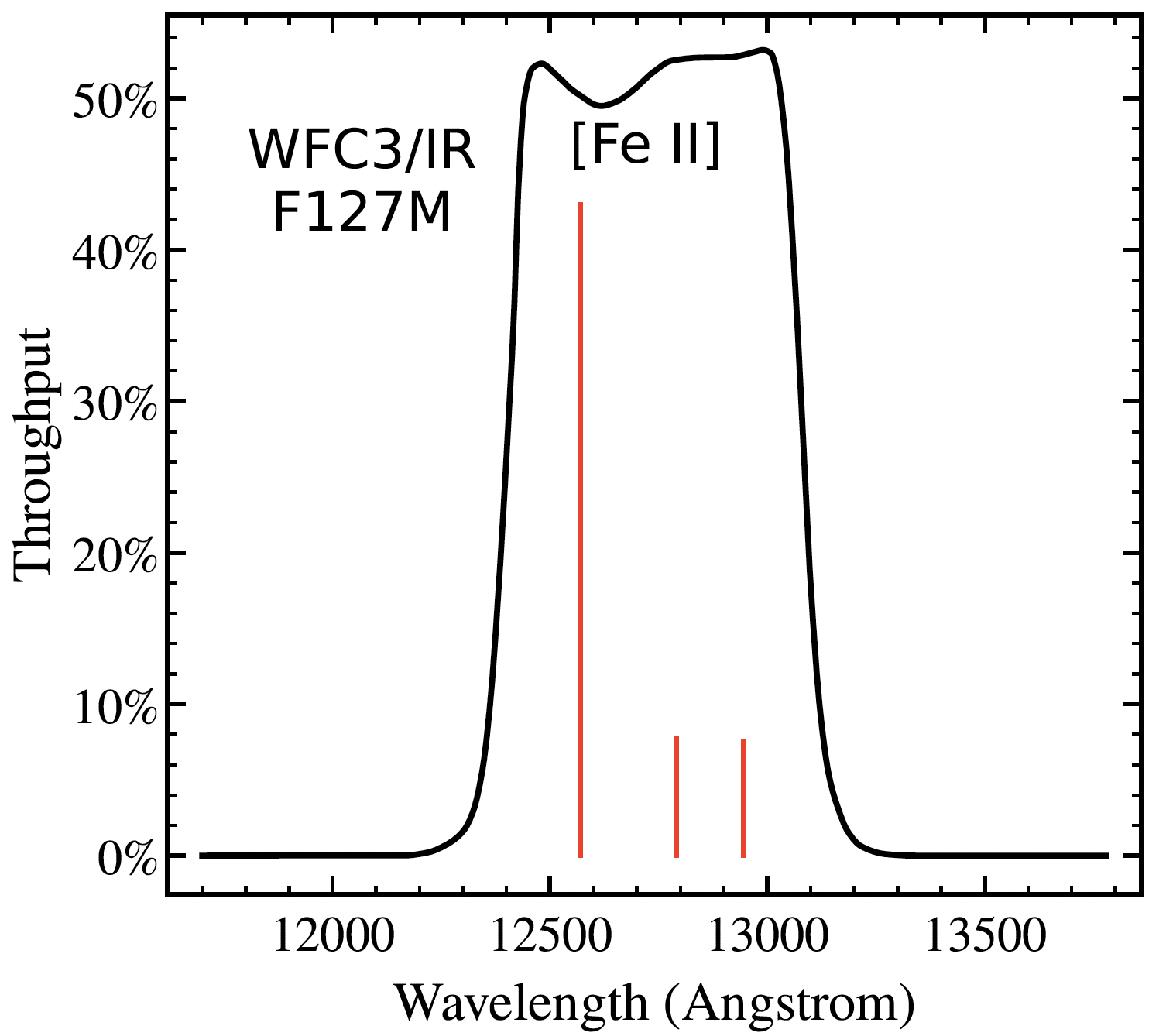}
\includegraphics[angle=0,width=5.4cm]{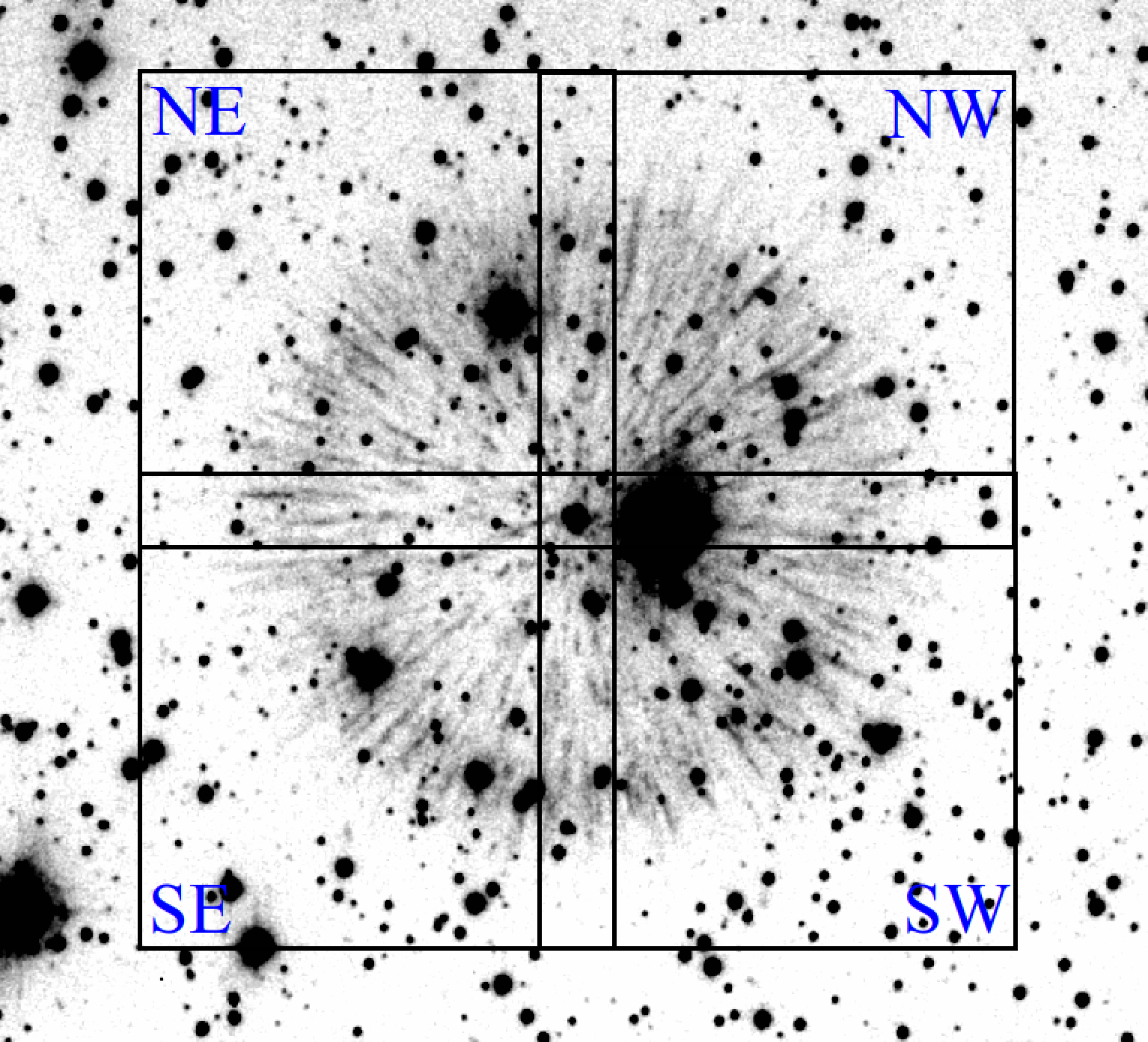}
\caption{Left and middle panels: Integrated system throughput 
for WFC3/IR filters F098M and F127M along with the 
wavelength locations of the near-infrared
[\ion{S}{3}] and [\ion{Fe}{2}] emission lines.
Right panel: Map of the four Pa 30 target positions. 
\label{plots} }
\end{center}
\end{figure*}
%%%%%%%*********************************

Much deeper and higher resolution 
[\ion{S}{2}]
images have recently been obtained using
the GMOS instrument on the 8-m Gemini North telescope 
that reveal the remnant to contain many hundreds more filaments, each of which appear composed 
of  
chains of small knots with apparent angular dimensions that remain 
approximately constant across the nebula \citep{Cunningham2026}. 
Gaussian fits
to the median profile of these knots 
indicate a characteristic FWHM of $0.68\pm 0.03''$.
Correcting for the images' point spread function (PSF) of $0.50''$, these filament knots  have 
a characteristic knot diameter of approximately $0.6''$, corresponding to a physical size of 
$\sim 2 \times 10^{16}$ cm.

The origin of the Pa~30's unusual optical appearance and numerous filaments is presently unclear. In light of the very high-velocity wind off its central star, explanations of wind-driven origins involving either mass ablation off dense ejecta knots \citep{Fesen2023}, or Rayleigh-Taylor (RT) instabilities in the remnant's shocked wind as it interacts with surrounding circumstellar medium \citep{Coughlin2026} have been proposed. Other proposed origins include Rayleigh-Taylor instabilities plus highly efficient nebula cooling via emission lines in Pa~30's metal-rich ejecta \citep{Duffel2024,Pikus2026}.

Here we present high-resolution near infrared Hubble Space Telescope (HST) images of Pa~30's nebula taken with the goal of investigating the sub-arcsecond structure of its filaments. Ground-based spectra that lead us to design our HST infrared imaging program, as well as details of our observing program and strategy are presented in $\S$2. This is followed by our HST imaging results and filament knot analyses in $\S3$. 
Comparisons with previous imaging are presented in $\S4$ along with a discussion of the significance of our results. Our main conclusions are given in $\S5$.

%______________________________
\section{Observations}
%______________________________

Low dispersion spectra were taken using the Large Binocular Telescope (LBT) of the Pa~30 nebula with the Multi-Object Dual Spectrograph 
(MODS; \citealt{pogge2012}) on 2020 October~10 and 2021 September~9. For each of these observations, a north-south 0.8~arcsec wide slit was placed on the central star, WD~J005311 with the main goal of studying the star's emission line variability \citep{Thomas2026}. In all, 67 individual images were combined for a total exposure time of 2.84~hours. Examples of the extracted knot spectra are shown in Figure~\ref{fig1}, and an image of the [\ion{S}{2}] emission along the slit is shown in the appendix (Figure~\ref{figA1}).

These spectra detected strong [\ion{S}{3}] 
$\lambda\lambda$9069,9531 emission lines in Pa~30 comparable to that of its [\ion{S}{2}] emission lines (Fig.\ 1). This plus the nebula's
relatively high extinction and the possible detection of some added emission from the near-IR [\ion{S}{2}] lines at  1.0287 - 1.0371 $\mu$m 
led us to select HST's WFC3/IR channel to image Pa~30's nebula.

The WFC3 IR channel uses a 1024 × 1024 Teledyne HgCdTe with a sensitive spectral range of 800 to 1700 nm. It has $1014 \times 1014$ rectangular pixels each measuring $0.135'' \times 0.121''$ and a field of view of $136'' \times 123''$. Because the WFC3/IR detector has several cosmetic defects including `blobs', poor sensitivity unbonded pixels, large areas of low sensitive pixels (the `wagon wheel'),  and
areas of dead pixels (the `death star'), we adopted several observing strategies
including integrating long enough to be limited by background emission and not read noise, dither exposures in large enough steps ($>$ 10 pixels) so that 
resolution can be restored to the diffraction limit and bad pixels and cosmic-ray impacts can be removed, and employing MULTIACCUM data blocks for readout noise suppression.

Despite these detector issues, WFC3/IR channel offered the promise of improved
filament detections compared to ACS/WFC or WFC3/UVIS images
due to its larger pixels, less extinction than in the optical (Pa~30: $E(B-V) \simeq0.88$ mag), and a higher throughput than the WFC3/UVIS using the F673N
[\ion{S}{2}] $\lambda\lambda$6716,6731
filter  (22\% -- 25 \%) versus the WFC3/IR F098M filter (42\% -- 52\%). Moreover, WFC3/IR F098M filter images will be sensitive to both [\ion{S}{3}] and [\ion{S}{2}] emission lines (Fig.\ 2) not just [\ion{S}{2}] as with the WFC3/UVIS F673N filter.
Pa 30's filaments are relatively faint with the brighter ones having a measured  [\ion{S}{2}] 
$\lambda\lambda$6716,6731 flux of $\simeq 2.5 \times 10^{-17}$ erg cm$^{-2}$ s$^{-1}$. Choosing the F098M filter would allow detection of knot [\ion{S}{3}] $\lambda$9531 emission which is similar in strength to the [\ion{S}{2}] $\lambda$6716 line, but not the other
[\ion{S}{3}] line at 9069 \AA,  which would be just outside of the F098M filter bandpass. 
In addition, images taken with the WFC3/IR F127M filter would be sensitive to the nebula's near infrared [\ion{Fe}{2}] emissions especially at 1.2570 $\mu$m (see Fig.~\ref{plots}).

%%%**********************************
%%%Figure 3: Raw Images
%%%%**********************************
\begin{figure*}[t]
\begin{center}
\includegraphics[angle=0,width=15.0cm]{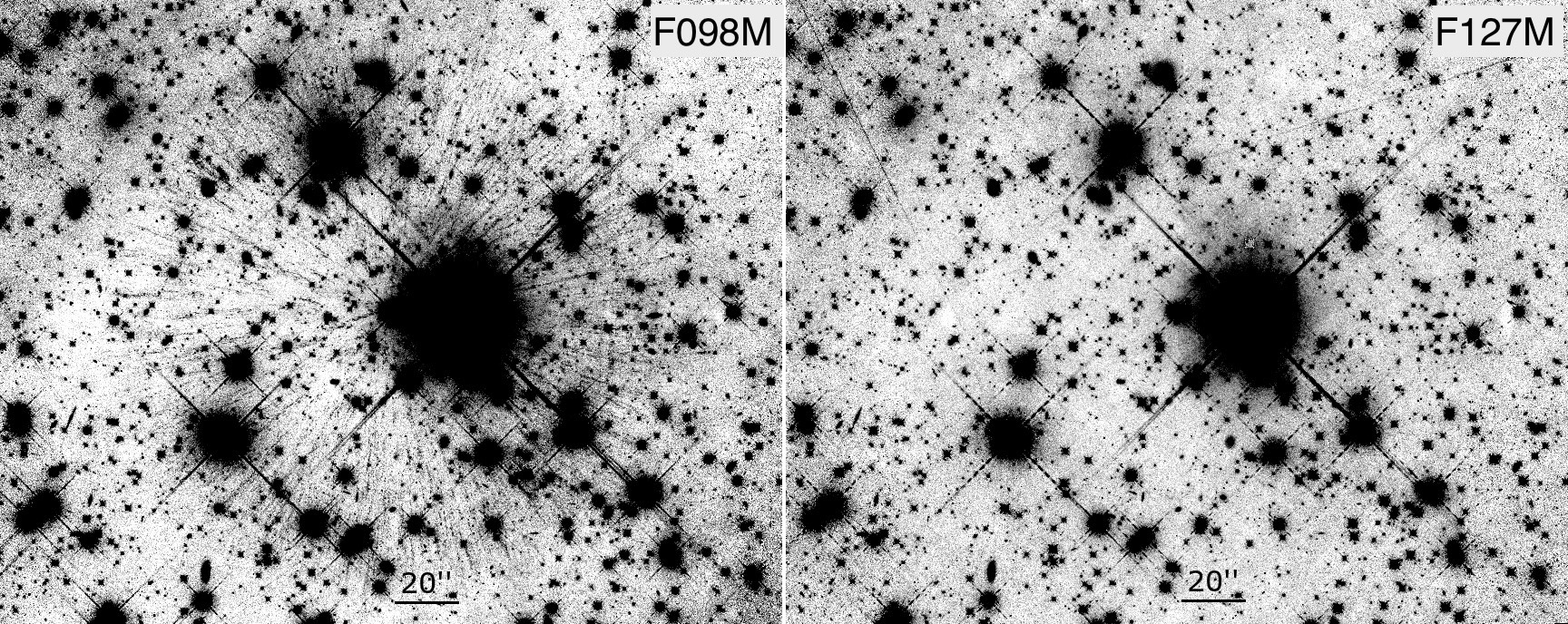} \\
\includegraphics[angle=0,width=15.0cm]{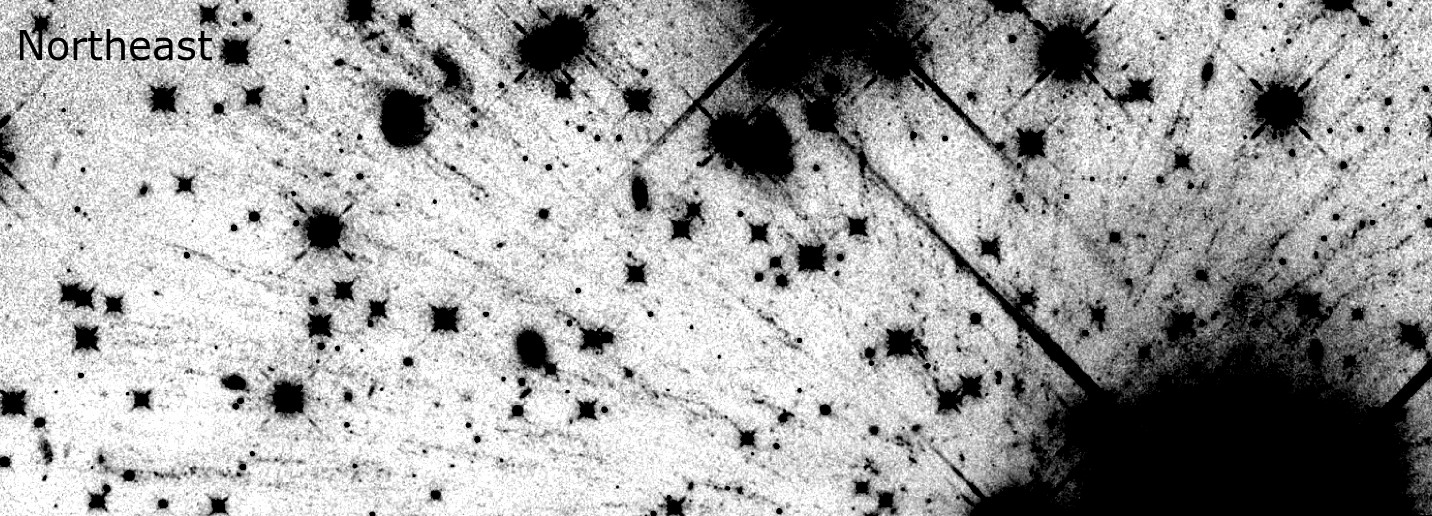} \\
\includegraphics[angle=0,width=15.0cm]{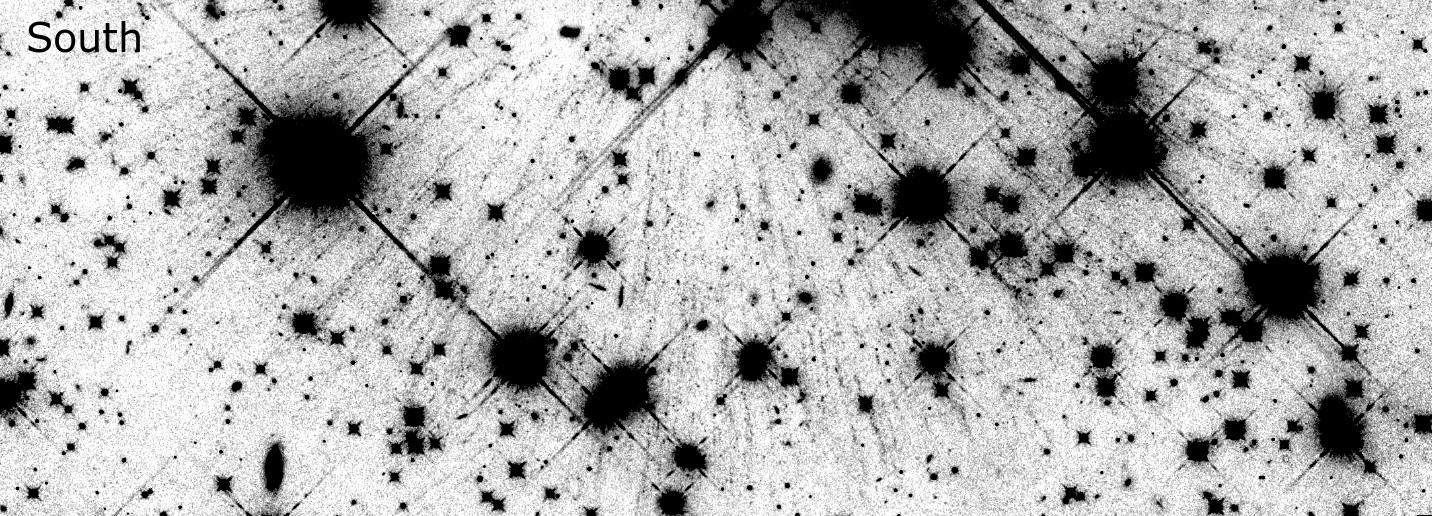} \\
\caption{Top panels: WFC3/IR F098M and 
F127M  filter images of Pa 30 corresponding to 
[\ion{S}{2}] + [\ion{S}{3}] and 
[\ion{Fe}{2}]. \\
Middle and lower panels:
Enlarged versions of the F098M image of the
remnant's northeastern and southern regions.
\label{fig3}
}
\end{center}
\end{figure*}
%*********************************

A series of WFC3/IR F098M and F127M images were taken (Program ID: 18127) on 2026 October 11 - 16 centered at the four regions shown 
in Figure 2 using four orbits for each region with exposures of $2 \times 2412$ s using F098M and $1 \times 4812$ s using F127M for each orbit. The data collection was  
accomplished in 16 orbits and in 4 visits.
Total exposures are about 19,300 s in each filter, from 32 and 16 combined F098M and F127M frames.
It turned out that no [\ion{Fe}{2}] emission was detected and thus this image was used as an off-band images to the F098M [\ion{S}{2}] + 
[\ion{S}{3}] image.

The data were taken in standard
MULTACCUM mode with aperture IR-FIX and SAMP-SEQ=SPARS50 and
NSAMP=11. The primary dither patten was a four point WFC3-IR-DITHER-BOX-MIN. This dither patten recovered more 
than 95\% of the WFC3/IR detector's bad pixels.
The resulting images are $4000 \times 4000$ mosaics drizzled at 
$0.09''$ per pixel and cover a field of view of $4.7' \times 4.3'$. 
Below we will often refer to the HST F098M and F127M images as [S~II+III] and background images, respectively.

Lastly, we downloaded the publicly available images of Pa~30 taken in June 2025 and recently reported on by
\citet{Cunningham2026} using the GMOS instrument on the Gemini North telescope. These images were helpful in the analysis of our HST image data.

%%**********************************
%%%Figure 4: Gemini vs HST Comparison
%%%%**********************************
\begin{figure*}[ht!]
\begin{center}
\includegraphics[angle=0,width=16.0cm]{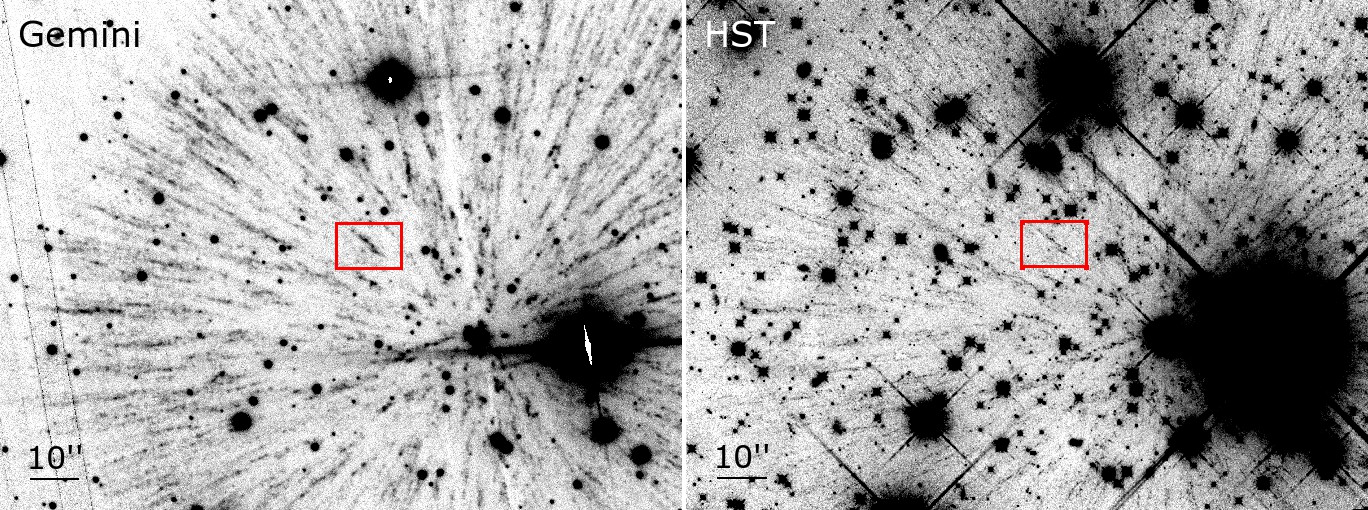} \\
\includegraphics[angle=0,width=16.0cm]{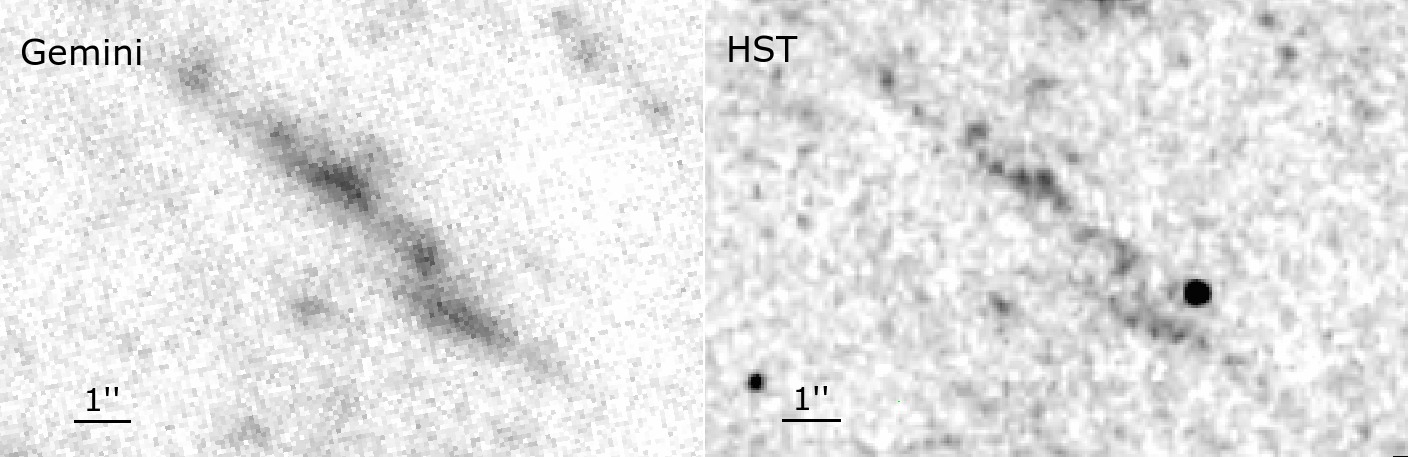} \\
\caption{Top panels compare Gemini GMOS [S~II] and HST WFC3/IR [S~II+III] images for Pa 30's northeastern region. 
The $13'' \times 9''$ red boxes mark a bright filament shown in enlargements in the bottom panels. This filament lies at the center of our `Field 1'. The image shows the HST image resolving this Pa~30 filament into a broken line of small knots. \label{fig4}
}
\end{center}
\end{figure*}
%%%%%%%*********************************

%%%%%%%%%%%%%%%%%%%%%%%%%%%%%%%%%%%%%%%%%%%%%%%%%%%%%%%%
%%%  Map of 15 Fields
%%%%%%%%%%%%%%%%%%%%%%%%%%%%%%%%%%%%%%%%%%%%%%%%%%%%%%%%%
\begin{figure*}[ht!]
\begin{center}
\includegraphics[angle=0,width=12.5cm]{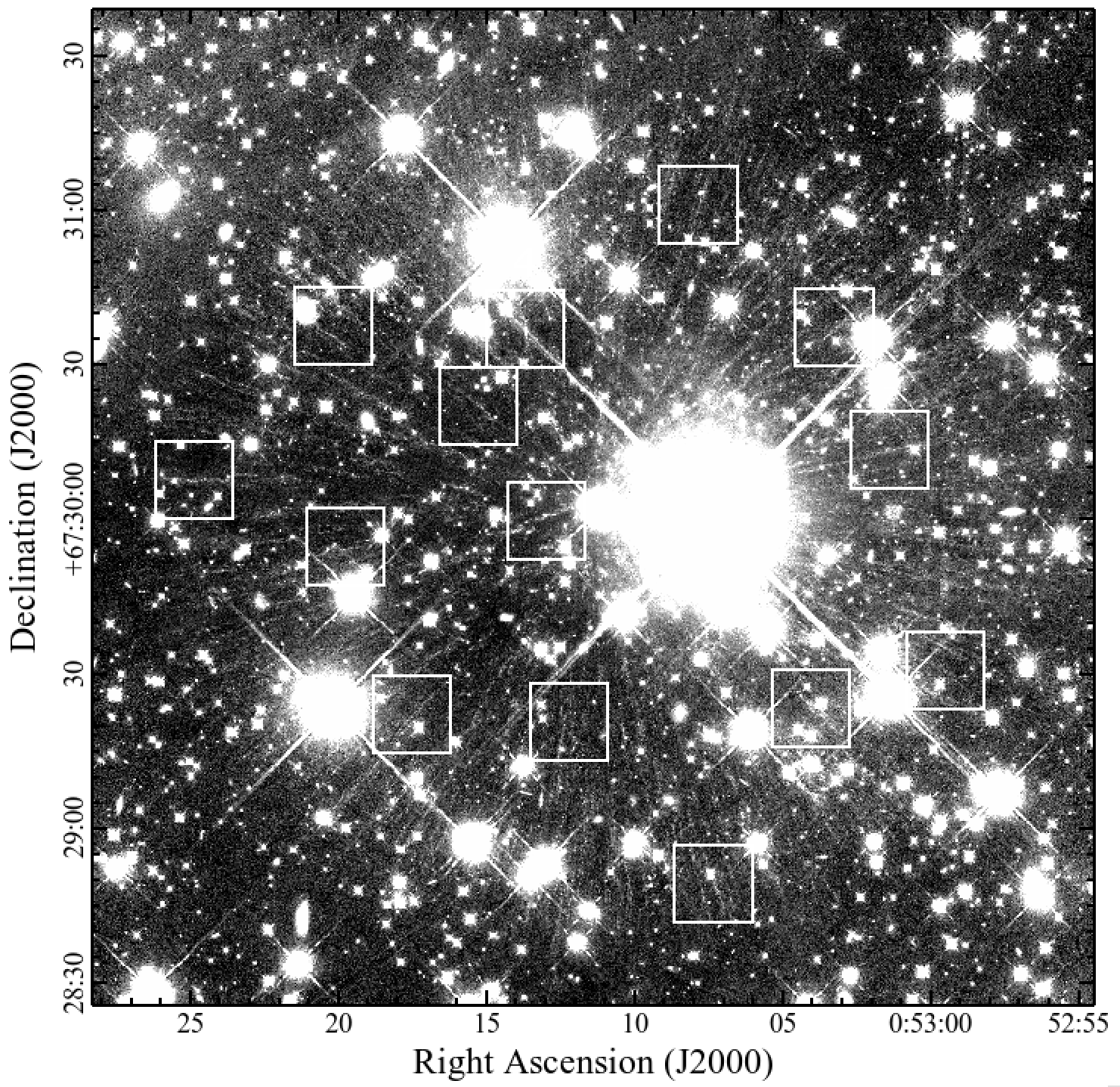}
\caption{Cropped HST WFC3/IR F098M image of Pa~30 showing the 
location of the 15 fields where filament knots were detected and measured.
Although the actual fields ranged in size from $11''$
to 21$''$, the target fields shown are all 
$15'' \times 15''$.
\label{Map} 
} 
\end{center}
\end{figure*}
%%%%%%%%%%%%%%%%%%%%%%%%%%%%%%%%%%%%%%%%%%%%%%%%%%%%%%%%%

%%%%%%%%%%%%%%%%%%%%%%%%%%%%%%%%%%%%%%%%%%%%%%%%%%%%%%%%%%%%%%%
%%% Table of Knot Results
\begin{deluxetable*}{ccccccc}[ht]
%\tiny
\centerwidetable
\tablecolumns{7}
\tablecaption{Knot Detections in 15 Pa 30 Filament Fields}
\tablewidth{0pt}
\tablehead{ 
\colhead{Field} & \multicolumn{2}{c}{\underline {Filament Centers (J2000) } } &
\colhead{All Knot}   & \colhead{Clean Knot} & 
\colhead{Median FWHM} & \colhead{Median FWHM}  \\
\colhead{ID} & \colhead{RA} & \colhead{Dec} &
\colhead{Detections} & \colhead{Detections} &   
\colhead{(observed)} & \colhead{(PSF Corrected)} 
} 
\startdata
 1 & 00:53:15.09 & +67:30:21.2  & 32 & 28 & $0.263''$ & $0.150''$ \\
 2 & 00:53:24.84 & +67:30:07.4  & 20 & 17  & $0.289''$ & $0.182''$ \\
 3 & 00:53:13.42 & +67:30:00.1  & 41 & 35  & $0.300''$ & $0.197''$ \\
 4 & 00:53:13.68 & +67:30:36.8  & 39 & 22  & $0.275''$ & $0.145''$ \\
 5 & 00:53:20.15 & +67:30:35.3  & 17 & 13  & $0.284''$ & $0.170''$ \\
 6 & 00:53:03.87 & +67:29:23.0  & 11 & 11  & $0.273''$ & $0.155''$ \\
 7 & 00:53:01.35 & +67:30:13.4  & 14 & 12  & $0.275''$ & $0.158''$ \\ 
 8 & 00:53:12.16 & +67:29:20.5  & 20 & 20  & $0.257''$ & $0.124''$ \\
 9 & 00:53:07.30 & +67:28:49.1  & 22 & 17  & $0.273''$ & $0.155''$ \\
10 & 00:53:19.73 & +67:29:54.5  & 25 & 15  & $0.277''$ & $0.162''$ \\
11 & 00:53:03.21 & +67:30:37.2  & 5  & 5   & $0.285''$ & $0.176''$ \\
12 & 00:53:07.81 & +67:31:00.9  & 14 & 9   & $0.292''$ & $0.186''$ \\
13 & 00:53:17.51 & +67:29:22.1  & 14 & 12  & $0.259''$ & $0.128''$ \\
14 & 00:52:59.66 & +67:29:35.9  & 9 & 7   & $0.258''$ & $0.126''$ \\
15 & 00:52:59.45 & +67:29:31.6  &  9 & 3   & $0.272''$ & $0.154''$ \\
\bf{ALL}&        &   & \bf{292} & \bf{226} &  \bf{0.273}$''$ & \bf{0.154}$''$ \\ 
\enddata
%\tablenotetext{a}{PSF = $0.225''$ was  for all knots. }
\end{deluxetable*}
%%%%%%%%%%%%%%%%%%%%%%%%%%%%%%%%%%%%%%%%%%%%%%%%%%%%%%%%%%%%%%%%

%-----------------------------------
\section{Results and Analysis}
%-----------------------------------

The resulting WFC3/IR F098M [S II+III] 
and F127M [\ion{Fe}{2}] background subtracted
images of Pa 30 are shown in the 
top two panels of Figure~\ref{fig3}. 
These images have been cropped to show a region 
approximately $240'' \times 190''$ centered on Pa~30's central star, IRAS00500+6713. 
While an uneven background is seen across both filter images, 
only the F098M image shows any appreciable nebular emission features. 

The bottom two panels in Figure~\ref{fig3} 
show enlargements of the F098M [S~II+III] image 
of Pa~30's northeastern and southern regions. 
Although dozens of filaments are detected, the 
S/N detection of the filaments is relatively low 
with typical peak knot S/N $\leq8$. 
However, the detection of knots in the remnant's brighter filaments 
is reasonably good and the images resolve the filaments into
narrow and highly discontinuous emission features. 

Figure 4 shows a comparison of Pa30's filaments seen in the Gemini North GMOS [\ion{S}{2}] image taken under excellent seeing conditions 
(PSF = 0.5$''$; \citealt{Cunningham2026}) and our F098M WFC3/IR [S~II+III] image. While a far greater number of fainter filaments are detected in the Gemini image, the HST image resolves those filaments into strings of small emission knots. 
According to the HST WFC3 handbook, the IR detector's intrinsic PSF is 0.126$''$ to 0.128$''$ for the wavelength of 900 to 1000 nm. The
effective PSF for our drizzled images is $0.225''$ because it includes the drizzling to $0.09''$ pixels, image-cut-out resampling,
and the smoothing used for knot detection.
Many knots in our HST images approach the $0.225''$ limit, meaning they are only marginally resolved.

A closer comparison of these images is shown in the lower panels of Figure 4 where we compare the Gemini [\ion{S}{2}] image (left panel;
\citealt{Cunningham2026}) with our F098M [\ion{S}{2}] + [\ion{S}{3}] image (right panel) for one of Pa~30's brighter filaments. What appears in the Gemini image as a fuzzy and lumpy filament is resolved into a discontinuous string of small knots whose sizes are near the resolution of the WFC3/IR images. Note: Because of both the F098M filter's much wider bandpass than that of the Gemini's 
narrow [\ion{S}{2}] filter ($\sim 1700$ \AA \ v.\ 90 \AA) and its 
infrared sensitivity, some faint stars appear in the
F098M images that are not detected in the Gemini image.

\subsection{Knot Detection and Analysis}

A visual inspection of the F098M image 
(PSF $\simeq0.225''$)
showed that Pa~30 filaments are made up of tight streams of small, sub-arcsecond emission knots
(see Fig.~\ref{fig4}). 
We measured the angular dimensions of
F098M emission knots manually in several of the remnant's brightest filaments. The filaments examined 
were concentrated largely in the eastern half of the nebula, both because there was less confusion with the bright 12 mag star near the nebula's center, and several bright filaments were located in the remnant's northern and eastern hemispheres.

The realization that all the nebula's filaments are composed 
of small emission knots lead us to develop a 
Python program to provide a more detailed analysis of Pa~30's filamentary emission. Guided by our selection of small regions with bright filaments, the program 
used co-registered Gemini GMOS [\ion{S}{2}] images to create masks to define the location and lengths of the filaments, confirm measured PSF-corrected FWHMs, median knot separations within a filament, and provide negative-image false-detection check.

In what follows are the results of our effort
to explore the size distribution of knots along several filaments in the remnant. 
Because the F127M filter image did not detect any remnant emission, we have used it as
an effective off-band image to help discriminate emission knots from field stars. 

We first chose seven fields in the remnant which contained at least one filament that was relatively bright.  This was later enlarged to 15 fields. The 15 fields chosen are indicated in Figure~\ref{Map}.
A filament knot detection mask was generated  
from the Gemini GMOS [\ion{S}{2}] image. This was done by first subtracting a $2.19''$ square local median background which was then smoothed, keeping only regions 
that are resolved and elongated (major-axis FWHM $\geq0.9''$) and not star-like (peak ($\leq30\sigma$). This rejected stars leaking through the GMOS [\ion{S}{2}] filter. 

The procedure to detect filament knots on the HST F098M image included the subtraction of a $1.32''$ running median then smoothed it with a Gaussian ($\sigma$ = $0.033''$). Masks were then created for bright stars with $>40\sigma$ and radius $\geq0.26''$.
The GMOS images were registered to HST using field stars, correcting offsets of up to $\simeq0.2''$ to a residual of $\leq$0.03$''$.

We defined a knot detection as 
a local maximum  
$\geq$ 4$\sigma$ in the smoothed F098M image within the GMOS
[\ion{S}{2}] filament mask, separated by 
$\geq0.087''$ from other maxima, was outside the masks around bright stars, and had a FWHM 
that is
at least half as wide as the WFC3/IR PSF which rejected hot pixels and cosmic-ray residues.

Because the F127M filter showed no line emission from the nebula, we used it to reject continuum sources. This included F127M detections with S/N $\geq$ 1.75 and an F127M-predicted stellar fraction $\geq$ 0.25. These were flagged and inspected by eye, as were detections within $2.5''$ of bright stars. Of 292 detections in our 15 filament fields, 226 were designed as `clean'  knots and were then used in our analysis. Of the remainder, 49 were designated as stars, stellar artifacts or stellar diffraction-spike features, plus another 17 that were ambiguous; e.g., weak F127M counterparts or proximity to stellar artifacts. These detections were thus excluded.

In making the Gemini GMOS filament mask, we first chose a
filament threshold signal of 3$\sigma$ but since several of our selected filaments are quite faint even in the Gemini image, this was revised down to $2\sigma$. This allowed for the detection
of more knots in fainter filaments.  

The F098M image's PSF was measured 
from field stars for Fields $1 –5$ ($0.216'' –0.234''$) and was fixed at $0.225''$ for Fields $6 – 15$, where automatic PSF star selection was unreliable due to  blended and/or saturated stars.
From an inspection of the results, this criteria lead larger knots and knot blends to being missed and hence uncounted. This was manually confirmed for several bright filaments. Our knot detection program also rejected knots that measured 
narrower than $0.5 \times$ the stellar derived PSF where knot PSF-corrected FWHM is equal to 
 (${\rm FWHM^{2} - {\rm PSF}^{2}})^{1/2}$.

\begin{figure}
\begin{center}
\includegraphics[angle=0,width=9.0cm]{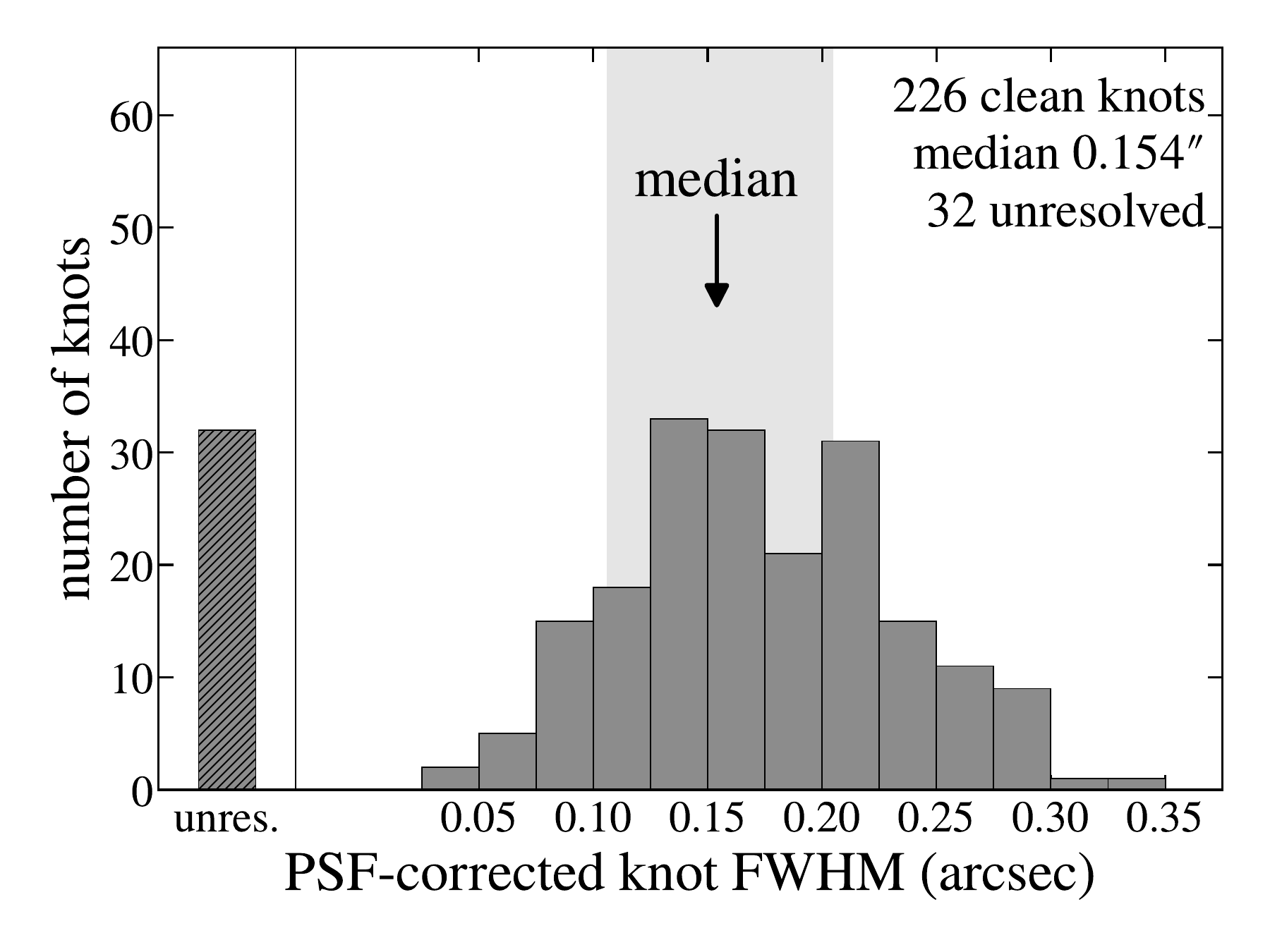} \\
\caption{Histogram of Pa 30 knot sizes}
\label{histo} 
\end{center}
\end{figure}

Table 1 lists details regarding filament knot 
detection in the 15 fields. The measurements for median FWHM (observed) and PSF corrected values 
used the measured PSF for Fields 1–5 and $0.225''$ 
for Fields 6–15.
Of the 226 clean knots detected in our 15 Pa~30 fields, 194 were larger than the instrumental PSF. The 226 knots were found to have a median PSF-corrected FWHM of
$0.154''$, and $0.168''$ with 32 unresolved knots removed,  i.e., with intrinsic FWHMs
$\lesssim0.1''$.
For the 194 resolved knots, 80\% fall between a FWHM of $0.095''$ and $0.25''$.
Figure~\ref{histo} shows the resulting histogram of the FWHMs  distribution for these knots.

%%%**********************************
%%%Figure 6: histogram
%%%%**********************************
%%%%%%%*********************************

%%%**********************************
%%%Figure knot profiles 15 Fields
%%%%**********************************
\begin{figure*}[ht]
\begin{center}
\includegraphics[angle=0,width=8.50cm]{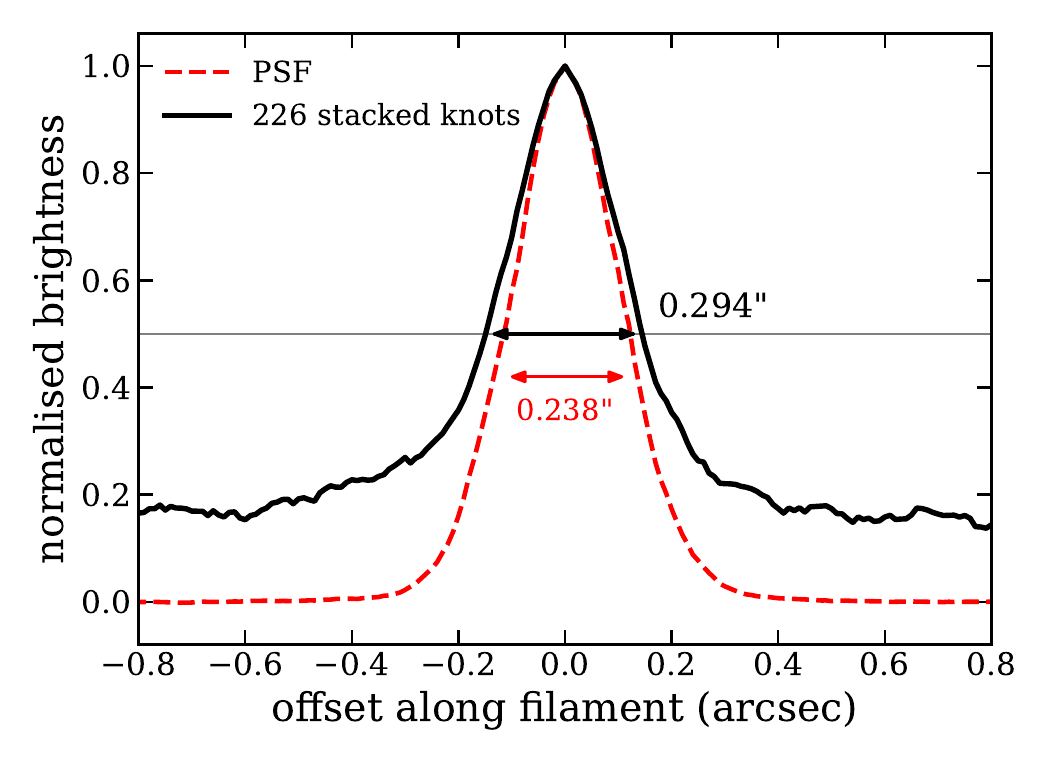}
\includegraphics[angle=0,width=8.5cm]{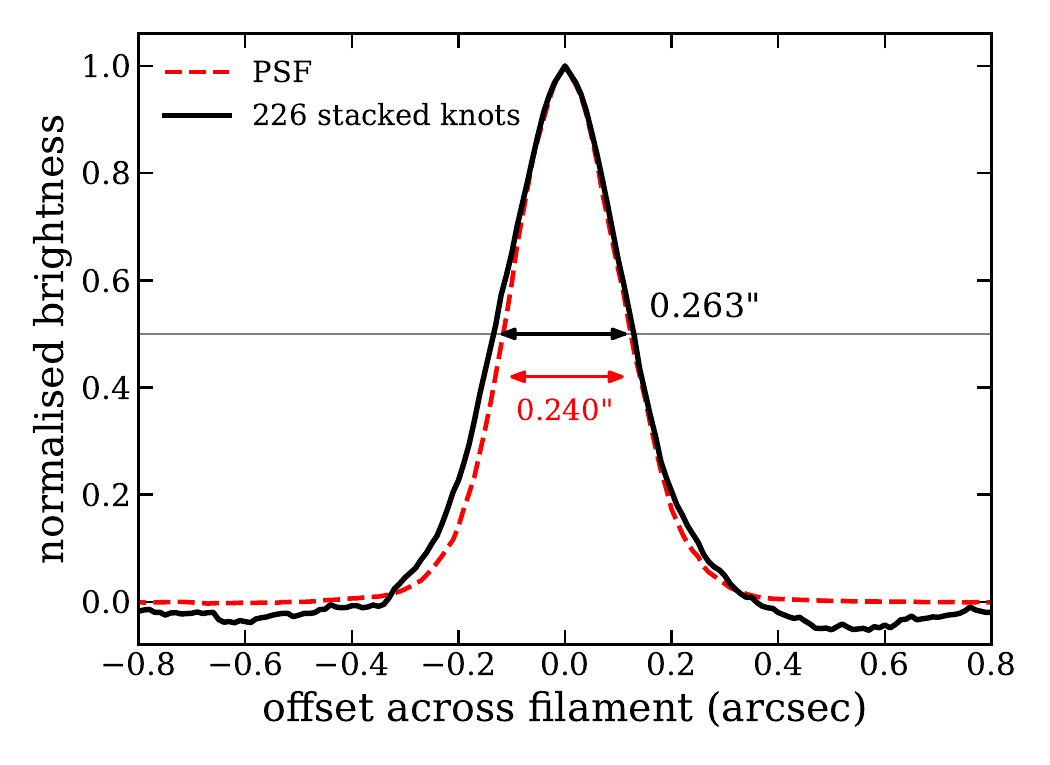}
\caption{Knot profiles for both along and across filaments. 
All knots are rotated to match the filament direction. The red dashed PSF represents a stack of 95 field stars processed identically. The along-filament wings seen are due to background filament emission.
\label{profiles} }
\end{center}
\end{figure*}
%%%%%%%*********************************

\subsection{Knot Stacking}

The majority of knots we detected had a peak 
emission of only $4-8 \sigma$ above background.
Consequently, a single knot's measured width
can be seriously affected by noise.
Individual knots sizes, therefore, can have large errors with the smallest ones are biased upwards. Moreover, knots often sit on faint filament emission and may have near neighbors. Thus looking at single knot measurements cannot easily discriminate a knot’s own light from its surroundings.

However, stacking all detected knots 
can average away this inherent noise.
Therefore, we combined all 226 knots, each
rotated to match its underlying filament 
orientation so all align horizontally.
Such a stacked profile  has an effective S/N
of over several dozen vs.\ 
4-8 for a single knot, thus  making knot
shapes measurable.

Figure~\ref{profiles} show our stacked knot emission profiles both along and across the filament directions. Knot widths  along filaments are clearly wider than the PSF; 
they are elongated, wider along the filament;
$0.294'' - 0.263''$.
Faint filament emission lies underneath some of the knots, which then shows up as wings in the plot for knot widths along the filament direction.
Because the knots were rotated to the filament direction before stacking, the along and across sizes can be measured separately. Stacking un-rotated knots would have blurred the two together. 
Moreover, if the underlying faint filament emission was roughly about the same size as the knots themselves, cross filament cuts would be expected to also show extended emission much like seen in the along filament cuts.

Knot stacking is independent of both individual knot measurements and elliptical fits. However, a check showed that they all agree and that the knots are resolved in these HST images at roughly $0.1'' - 0.2''$ and are
elongated along the filaments.

We note that stacked stellar profiles have half-maximum widths $\simeq0.24''$, a bit larger than the $0.225''$
adopted for our knot PSF correction. This is because our star stacks are centered on each object’s brightest pixel and combine stars from all 15 fields. Because of this knot stacks are affected in the same way and are therefore compared with these star stacks.

Although stacking of knots is helpful in that it gives a good estimate regarding the typical
knot size, the histogram shown in Figure~\ref{histo} gives the actual spread in sizes.
The gray band is the interquartile range ($0.106'' – 0.205''$),  and that `unres.' represents the 32 unresolved knots.
and that the median marks $0.154''$.
But stacking blurs the resulting knot sizes slightly due to small errors in peak knot position and in estimating filament direction. 

%%**********************************
%%% Big Figure of knot profiles
%%%%**********************************
\begin{figure*}
\begin{center}
\includegraphics[angle=0,width=17.50cm]{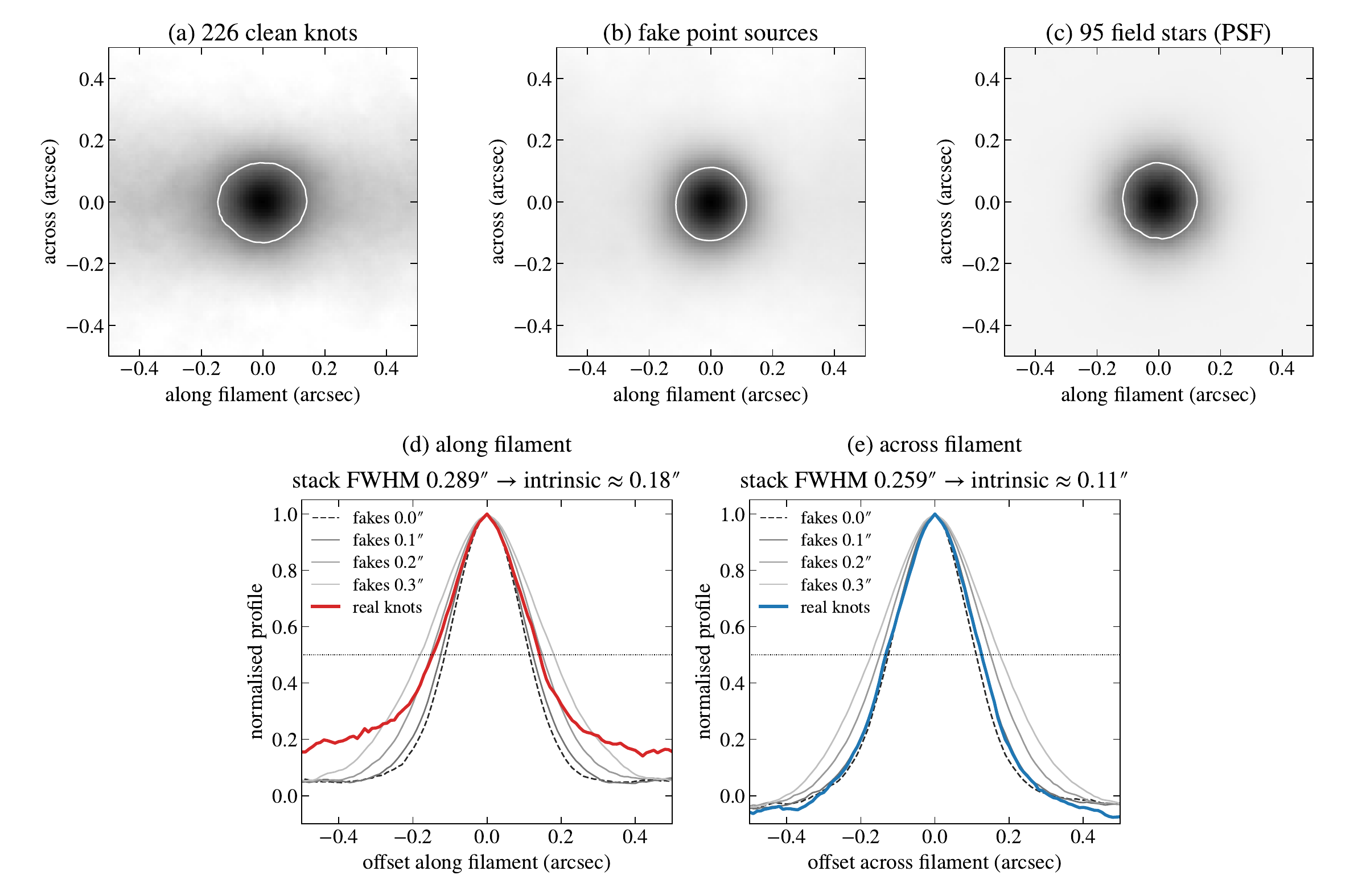}
\caption{Stacked F098M images and profiles of the Pa~30 knots. (a) Image cut-outs of the 226 clean knots, (b) image of stacked simulated point sources placed with the S/N of the real knots, (c) point-spread function of 95 field stars. (The white contours mark half maximum.) (d) and (e) Plots of cuts through the knot stack along and across real filament knots compared with stacks of simulated knots of known FWHMs. See text for details.
\label{fake}
}
\end{center}
\end{figure*}
%%%%%%%*********************************

%%%%%%%-----------------------------------

%------------------------------
\subsection{Artificial Knot Tests}
%---------------------------------

In order to test our knot detection method for both accuracy and completeness, we ran different sets of artificial or `fake' knot tests through our knot detection program. 
Each was built of measured stellar
PSF values and used intrinsic fake knot sizes 
placed on the nebula's filaments.

Test 1: Round Knots. To calibrate our 15 field knots detection percentage, we placed 31,494 fake knots with FWHM values of $0.00'', 0.05'', 0.10'', 0.15'', 0.20'', 0.25''$, and $0.30''$. The program recovered 27,371 of these or roughly 87\%. 

Test 2: Elongated Knots. In order to assess whether our along/across knot method recovers knot elongations, we placed 5,759 knots with 
$0.22'' \times 0.11'', 0.11'' \times 0.22'', 0.30'' \times 0.10''$ in both along and across
filament directions. The program recovered 5,084 of these knots or 88\%. 

In each of the 15 filament fields, up to 12 simulated knots 
were placed at random positions inside the GMOS filament mask, 
with the following constraints:
at least $0.6''$ from any actual real knot detection, 
at least $1''$ from each other, and 
at least $0.65''$ from the field image edge.
Each fake was rotated to the local filament direction and given a random shift of less than one pixel. And each simulated peak S/N was drawn at random from the real clean knots’ values (median 5.2, range 4.0–15.8). So the simulated knots have the same brightness distribution as the real ones, and they sit on real filaments with real noise.
For the round knot set, we ran 25 rounds × 7 sizes × 15 fields × 12 knots or about 2,100 per field and about 4,500 per knot size.
Recovered knots were measured exactly like real ones, that is half-maximum-area FWHM with PSF correction and the along/across elliptical fit.

The results for these artificial knot test are given in Table 2.
Knot recovery is high and nearly constant (85\% - 89\%),
falling only for the largest knots whose flux is more spread out.
Artificial knots processed through the same pipeline as done for real knots
show that the method recovers sizes accurately near the
median knot size. This indicates our $0.154''$ as the median size is reliable;
input $0.15''$: recovered $0.150''$.

The chief limitations to these tests are that the simulated knots are simple Gaussian blobs, the HST images underwent smoothing, and 
our method compresses the size range with the effect being overestimating the smallest knots and underestimating the largest ones. In addition, the recovered sizes in Fields 8--15 are systematically $\sim 0.02''$ larger than in Fields 1--7 for the same input size. In addition, individual fields can be offset by up to $\sim 0.04''$.

%%%%%%%%%%%%%%%%%%%%%%%%%%%%%%%%%%%%%%%%%%%%%%%%%%%%%%%%%%%%%%%
%%% Table Knot Fake Test Results %%%
\begin{deluxetable}{cccc}[ht]
\tablecolumns{4}
\tablecaption{Artificial Knot Test Results}
\tablewidth{0pt}
\tablehead{
\colhead{True} & \colhead{Recovered}  & \colhead{Measured} & \colhead{Measured} \\
\colhead{FWHM} & \colhead{Percentage} & \colhead{Median}   & \colhead{Unresolved}
} 
\startdata 
0.00'' & 88\% & 0.059'' & 43\%  \\
0.05'' & 89\% & 0.080'' & 37\%  \\
0.10'' & 89\% & 0.111'' & 30\%  \\
0.15'' & 89\% & 0.150'' & 20\%  \\
0.20'' & 88\% & 0.181'' & 12\%  \\
0.25'' & 85\% & 0.204'' & ~7\%  \\
0.30'' & 81\% & 0.227'' & ~5\%  \\
\enddata
%\tablenotetext{a}{Assuming d = 2.3 kpc.}
\end{deluxetable}
%%%%%%%%%%%%%%%%%%%%%%%%%%%%%%%%%%%%%%%%%%%%%%%%%%%%%%%%%%%%%%%%

Comparisons of real knot and simulated results
are shown in images and plots in Figure 8.
Panel (a) is a stack of F098M image cut outs of our 226 clean knots. These images have been rotated so that the filament direction, measured on the GMOS image, is horizontal. Faint underlying emission can be seen running through the knot stack.
Panel (b) is a stack of simulated point sources placed with the same S/N as our real knots and stacked in the same way. Panel (c) shows the point-spread function of 95 field stars in the F089M image, again rotated to the filament direction. White contours in these three panels mark half maximum values. 

Plots in panels (d) and (e) are cuts through the real knot stack along and across filaments compared with stacks of simulated knots of known FWHM 0.0, 0.1, 0.2 and 0.3 arcseconds. Intrinsic knot sizes  were found by interpolating the measured half-maximum width among the simulated stacks. As can be seen in the
two profile plots, the real knot curve measured along filaments of $0.289''$ falls between the
$0.1''$ and $0.2''$ fake knot curves giving $0.18''$, while across
filaments the real knot value of $0.259''$ falls just above the $0.1''$ fake curve
giving $0.11''$. The extended profile wings in (d) come from addition filament emission detected around each knot. These plots show that the knots are resolved and elongated along the filaments, with intrinsic FWHM 
$\simeq0.18''$ along and $\simeq0.11''$ across the knots, that is 410 AU and 250 AU at a distance of 2.3 kpc.

In these stack plots, stellar derived instrumental PSFs 
of $0.238''$ and $0.240''$, not the $0.225''$ used earlier. The value $0.225''$  was used to correct individual knot sizes and was measured in fields 1 - 5 ($0.216''$ - $0.234''$)
and was fixed for fields 6 - 15. The values of  $0.229''$ and $0.239''$ 
are widths of a slice through our stacked stars, made the same way as we did for the knot stack. An issue that make them these stellar PSFs slightly wider is that each star was centered on its brightest pixel and not its true center which slightly blurs the star
stack, and and the knot stack is blurred by the same amount.

Keeping the stacked-star PSF in these plots is the fairer comparison, because it went through exactly the same processing as the knots. And the difference barely changes the resulting values for true knot sizes. Using the stacked PSF values we get 
$0.171''$ and $0.129''$ versus $0.180''$ and $0.136''$ for along and across directional
measurements.

There are several important aspects to keep in mind about these plots. First, the faint horizontal band seen in panel (a) is ambient filament emission some knots happen to sit on, not the knots themselves. Every knot was rotated so its filament runs horizontally before stacking. So any filamentary emission that runs through each knot therefore lines up and adds into a horizontal band. This emission consists of both background filament emission which is roughly 15\% - 20\% of the peak of the stacked knot plot and is seen as the flat wings in panel (d), plus some emission from neighboring knots. The median spacing of knots is $\approx0.6''$. Consequently,
a neighboring knot along a filament often falls near the edge of the knot cut-out image which then will add to emission along the horizontal axis.

A perfectly round knot sitting on a filament would produce exactly the same horizontal band, so the band conveys nothing about the knot’s true shape. What it does show is that many knots are embedded in filament emission. However, evidence that the knots are truly elongated and not round comes from the knot cores, near their half maximum, where the any filament emission contributes little.

In the lower panels of Figure~8, the half-maximum width for the stack of knots made from cuts along the filaments is $0.289''$ whereas is $0.259''$ cut across the knots. Thus, the white contour seen in panel (a) is a little wider than tall. When we fitted each knot 
individually, 75\% of knots were found to be longer along the filament, whereas only 54\%
of simulated knots placed on the same filaments showed any elongation.
Moreover, tests using elongated simulated knots, e.g., $0.22'' \times 0.11''$ in shape,
show the same measured sizes as real knots.

What the horizontal band of filament emission does do is add a little light around the knots in the along direction which can slightly inflate our along-filament width results. 
But our individual knot fits include a local background, and the simulated knots sat on real filaments, so we believe this effect is largely accounted for.

Two final notes: These results assume true knots have Gaussian profiles which may not be very realistic given their true physical size being many hundreds of AUs.
Also the
stars used in this analysis are much brighter (S/N = 15 - 400) than the knots (SN = 4 - 8) which led to the star stack being far less noisy. 

%---------------------------
\subsection{Knot Fluxes}
%---------------------------
We have used our F098M image to estimate a knot brightness plus the total flux detected in the near infrared emission lines of [\ion{S}{3}]
and [\ion{S}{2}]. We measured each knot brightness within a radius of $0.26''$ and then subtracted the detected local flux at $0.4'' - 0.6''$ away, which could include some background filament light. 

This value was then corrected for light outside our chosen radius. According the WCFC3 handbook, 
only about 82\% of the light from a point source will fall inside a radius of $0.26''$ 
and slightly less for a knot of $0.22'' \times 0.12''$, giving 0.75 in total
when using the F098M filter. Since our measured Pa~30 knots are not point sources, the loss will be slightly greater, so we multiplied  the measured values by 1/0.75.

Converting count rates to flux numbers was done using STSCI's calibration tables which for the
F098M filter has a pivot wavelength of 9864 \AA \
and a PHOTFLAM of $6.05 \times 10^{-20}$ erg cm$^{-2}$ s$^{-1}$ per 
e$^{-}$ and a bandwidth of 157 nm. For emission lines, 
flux = count rate $\times$ PHOTFAM $\times$ bandwidth which gives $9.5 \times 10^{-17}$
erg s$^{-1}$ cm$^{-2}$ per e$^{-}$/s. 

Our flux measurements are listed in Table 3. 
A typical clean knot has a flux of about $3 \times 10^{-17}$
erg s$^{-1}$ cm$^{-2}$ in the WFC3/IR F098M bandpass.
This corresponds to a luminosity of about
$2 \times 10^{28}$ erg s$^{-1}$ at 2.3 kpc.
This typical knot flux number is for the combination of [\ion{S}{2}] and [\ion{S}{3}] lines emissions.

%%%%%%%%%%%%%%%%%%%%%%%%%%%%%%%%%%%%%%%%%%%%%%%%%%%%%%%%%%%%%%%
%%% Table Knot Fluxes %%%
\begin{deluxetable}{lccc}[ht]
\tablecolumns{4}
\tablecaption{Pa 30 Knot Flux Measurements}
\tablewidth{0pt}
\tablehead{ 
\colhead{Clean Knots} & \colhead{Count Rate} & \colhead{Flux} & \colhead{Luminosity\tablenotemark{a}} \\
\colhead{(N = 226)}  & \colhead{(e$^{-}$/s)} & \colhead{(erg/s/cm$^2$)}  & \colhead {(erg/s)}  } 
\startdata 
10$^{\rm th}$ percentile  & 0.18  & $1.7 \times 10^{-17}$ & $1.0 \times 10^{28}$  \\ 
Median                    & 0.32  & $3.0 \times 10^{-17}$ & $1.9 \times 10^{28}$  \\
90$^{\rm th}$ percentile      & 0.50  & $4.7 \times 10^{-17}$ & $3.0 \times 10^{28}$  \\
Sum of 226 Knots      &       & $7.3 \times 10^{-15}$ & $4.6 \times 10^{30}$  \\
\enddata
\tablenotetext{a}{Assuming d = 2.3 kpc.}
\end{deluxetable}
%%%%%%%%%%%%%%%%%%%%%%%%%%%%%%%%%%%%%%%%%%%%%%%%%%%%%%%%%%%%%%%%

It should be noted that these luminosity values are not corrected for extinction. With the published reddening of around 
$E(B-V)$ $\simeq 0.88$ mag
\citep{Lykou2023}, the extinction near 1 $\mu$m is about 1 magnitude. That would raise the luminosities by a factor of roughly $2–2.5$, up to about
$5 \times 10^{28}$ erg s$^{-1}$
for a typical knot.

Our knot flux estimates make it clear that Pa~30's knots are relatively faint emission features.
It is this faintness that lead to our
poor S/N filament detections despite fairly long exposures. Future observations with JWST could greatly improve both flux values as well as better determine knot sizes and deviations from spherical shapes.

%%%%%%%%%%%%%%%%%%%%%%%%%%%%%%%%%
\begin{deluxetable*}{ccccccccccl}[ht]
\centering
%\tablewidth{0.5pt}
\tablecaption{Emission Line Fluxes\tablenotemark{a,b}   \label{spec}}
\tablehead{
\colhead{Knot} & \colhead{$\theta^c$} & \colhead{RV} & 
\colhead{[\ion{O}{3}]} & \colhead{[\ion{S}{2}]} & \colhead{[\ion{S}{2}]} & \colhead{[\ion{Ar}{3}]} & \colhead{[\ion{S}{3}]} & \colhead{[\ion{S}{3}]} & \colhead{$\lambda$6716/$\lambda$6731} & \colhead{$n_{e}^d$} \\
\colhead{ID} & \colhead{(arcsec)} & \colhead{(km~s$^{-1}$)} & \colhead{$\lambda$5007} & 
\colhead{$\lambda$6716} & 
\colhead{$\lambda$6731} & \colhead{$\lambda$7136} & \colhead{$\lambda$9069} & \colhead{$\lambda$9531}  &
\colhead{(ratio)} & \colhead{(cm$^{-3}$)} }
\startdata
A1 & $+13.4$ & $+864(5)$ & \nodata  & 6.9(2)  & 5.9(2)  & 1.8(3) & 5.4(3) & 9.4(3)  & 1.18(4)  & 325$^{+68}_{-62}$    \\
A2 & $+13.4$ & $-882(5)$ & \nodata  & 4.2(3)  & 3.2(3)  & 0.8(3) & 1.9(5) & 1.9(5)  & 1.32(11)  & 134$^{+144}_{-112}$   \\
B  & $-12.2$ & $+843(5)$ & 8.4(6)   & 13.3(1) & 10.3(1) & 2.7(2) & 7.0(3) & 12.8(4) & 1.30(2)  & 158$^{+25}_{-24}$  \\
C  & $-31.8$ & $-950(5)$ & 8.7(6)  & 7.2(2)  & 5.5(2)  & 1.8(4) & 5.3(3) & 9.1(3)  & 1.32(5)  & 134$^{+61}_{-54}$   \\
D  & $+17.3$ & $+751(5)$ & \nodata  & 5.3(2)  & 4.1(3)  & 1.7(3) & 3.2(3) & 5.0(3)  & 1.30(8)  & 158$^{+106}_{-88}$   \\
E  & $-73.3$ & $-5(5)$ & \nodata  & 3.0(4)  & 2.9(4)  & \nodata & \nodata & \nodata  & 1.05(19)  & 580$^{+660}_{-350}$   \\
F  & $+71.7$ & $-349(5)$ & \nodata  & 2.1(4)  & 1.5(4)  & \nodata & \nodata & \nodata  & 1.40(25)  & 50$^{+320}_{-43}$   \\
\enddata
\tablenotetext{a}{See Appendix A for Knot ID locations.}
\tablenotetext{b}{Observed flux in 10$^{-17}$ erg~cm$^{-2}$~s$^{-1}$; not corrected for extinction.}
\tablenotetext{c}{North ($+$) and south ($-$) angular distance from stellar remnant WD~J005311.}
\tablenotetext{d}{Electron density assuming $T_e=10000$ K}
%\tablenotetext{}{Uncertainties are indicated by the final digit shown in parentheses.}
\end{deluxetable*}
%%%%%%%%%%%%%%%%%%%%%%%%%%%%%%%%%%%%%%%

The WFC3/IR F127M image, whose bandpass includes [\ion{Fe}{2}] 1.257 $\mu$m, showed no emission from the nebula. STScI lists a pivot of 1.2740 $\mu$m and 
a PHOTFLAM $9.33 \times 10^{-20}$ erg cm$^{-2}$ s$^{-1}$ per e$^{-}$, with a bandwidth of 68.8 nm and peak throughput 0.54. 
That gives $6.4 \times 10^{-17}$ erg s$^{-1}$ cm$^{-2}$
per e$^{-}$/sec. [\ion{Fe}{2}] 1.2567 $\mu$m lies well inside F127M bandpass, even when shifted by $\pm1400$ km s$^{-1}$.

Summed over all filament pixels in our 15 fields, we find the F127M flux to be $-0.002 \pm 0.016$ times the F098M ([S II]+[S III]) flux; at the positions of the 226 clean knots it is $0.031 \pm 0.008$, which may reflect faint continuum contamination. We therefore estimate a $3\sigma$ upper limit on [\ion{Fe}{2}] 1.257 µm emission of $\lesssim 5\%$ of the combined [S II]+[S III] flux in F098M, i.e., $\lesssim 1.5 \times 10^{-18}$ erg s$^{-1}$ cm$^{-2}$ for a typical knot. Thus we conclude
that the nebula is less than about 5\% in [\ion{Fe}{2}] emissions as bright as detected in the F098M filter.

We note that our flux estimates are only good to about $\pm$30\% due to several factors. One is that this conversion assumes that each line sits near the filter's
transmission peak. While this is true for the
[\ion{S}{3}] $\lambda$9531 and
[\ion{S}{2}] lines at 1.03 $\mu$m, it is not true
for the [\ion{S}{3}] $\lambda$9069  line which sits at the blue edge of the filter's bandpass.
Another issue is close neighboring knots plus the faint emission in the structure of the filaments which together can add scatter at the 
10\% - 20\%. Lastly, these numbers are not corrected for extinction. 
At $E(B - V) = 0.88$ mag, extinction is about 0.25 mag lower at 1.26 $\mu$m than at 0.99 $\mu$m, so the extinction-corrected limit for the [\ion{Fe}{2}] 
1.2570 $\mu$m line
would be roughly 
20\% tighter so about 4\%.

Finally, we can used these flux values to get a very rough estimate of the mass of sulfur in these knots.  
Correcting the total F098M flux of the 226 clean knots for extinction 
($E(B-V)$) = 0.88 mag) which is
$1.2 - 1.4 \times 10^{31}$
erg s$^{-1}$, 
and assuming a knot ionization level where the S$^{+}$/S$^{+2}$ ratio is 
near unity, 
and adopting T$_{\rm e}$ $= 10^{4}$ K and n$_{\rm e}$ $= 200$ cm$^{-3}$ (see $\S3.5$), 
we estimate a combined S$^{+}$ + S$^{+2}$ mass of 
$\sim 2 \times 10^{-9}$ M$_{\odot}$ per knot
or 
$\sim5 \times 10^{-7}$ M$_{\odot}$ for all 226 knots in our sample. 

This estimate has an uncertainty of about
an order of magnitude in either direction owing to the unknown 
[\ion{S}{2}]/[\ion{S}{3}] contributions to the F098M flux, and knot electron density and
temperature. 
If the [\ion{S}{2}] lines dominated the detected
F098M flux, then the mass estimate increases but becomes less if 
[\ion{S}{3}] emission was the main emission detected.
Lastly, if there are 1000 to 2000 knots in the whole nebula similar to the 226 we measured, then the nebula's total sulfur mass is larger by a factor of some 5 - 10 to around $2 - 4 \times 10^{-6}$ M$_{\odot}$.

\subsection{Knot Line Emissions and Electron Densities}

There has not been a published optical spectrum of Pa~30 knots showing the relative strengths of the nebula's optical emissions. 
Considering the compact nature of Pa~30's knots, 
the range of electron densities in the knots would be helpful in modeling their formation. 

\citet{Fesen2023} reported
a range for the density sensitive 
[\ion{S}{2}] $\lambda$6716/$\lambda$6731 flux ratio of
between 1.5 and 0.9, indicating electron densities between $<$100 and $\sim$700 cm$^{-3}$
with an mean ratio of  1.30, implying an
electron density of around 100 - 200 cm$^{-3}$ assuming an electron temperature of $10^{4}$K. 
Ratios around 1.3 are  also consistent with knot spectra shown in \citet{Cunningham2024}.

%%%**********************************
%%%Figure 9: Sequence
%%%%**********************************
\begin{figure*}[ht]
\begin{center}
\includegraphics[angle=0,width=16.0cm]{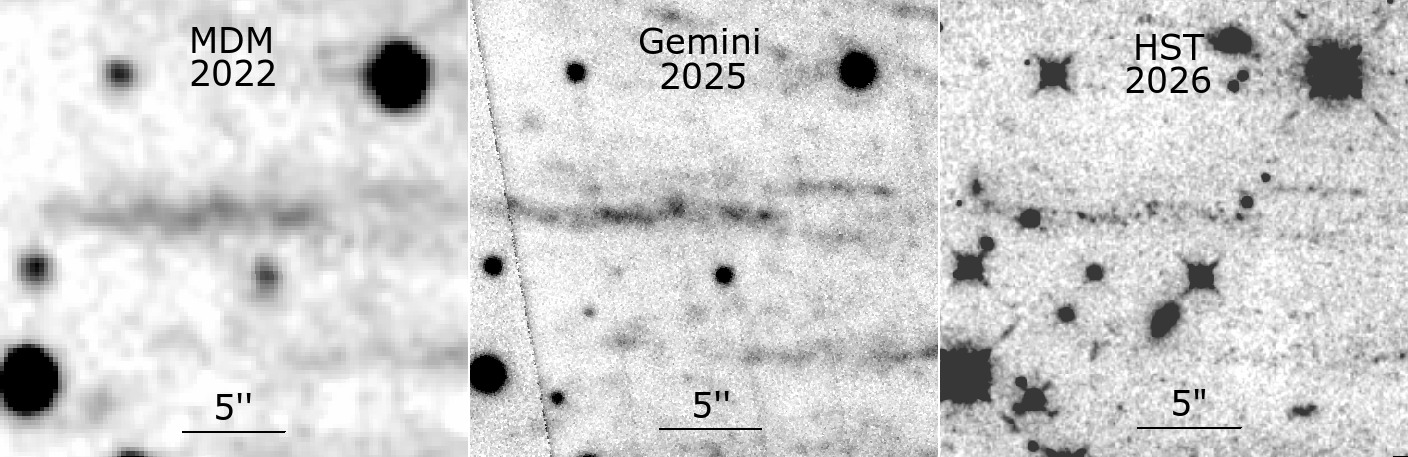} \\
\caption{Sequence of increasing resolution of Pa 30 [\ion{S}{2}] filaments; 
MDM (1.1$''$), Gemini ($0.50''$),
HST ($0.22''$). \label{res}
}
\end{center}
\end{figure*}
%%%%%%%%%%%%%%%%%%%%%%%%%%%%%%%%%%%

We can use the low-dispersion LBT spectra described in $\S2$ and shown in Figure 1 to measure knot line emissions. Spectra of seven individual knots (see Appendix A) were extracted using packages in PyRAF/IRAF\footnote{Interactive Reduction and Analysis Facility %\citep[IRAF;][]{tody86,tody93} 
was distributed by the National Optical Astronomy Observatories, which are operated by the Association of Universities for Research in Astronomy, Inc., under cooperative agreement with the National Science Foundation.}. The trace of WD~J005311 was used to guide the aperture over the location of the knots. These spectra were wavelength calibrated using neon and argon lamps and the flux calibrated from spectra of spectrophotometric standards obtained during the observing runs.

The relative fluxes for emission lines that were clearly detected in the brighter knots that intersected the slits were measured and are listed in  Table~4. While [\ion{O}{3}] $\lambda$5007 is weak in most knots, it is detected in the brighter ones. Correcting for extinction,
the $\lambda$5007 line would be roughly 1.5 times brighter than the [\ion{S}{2}] $\lambda$6716 line in those knots.
This is consistent with the finding of \citet{Cunningham2026} that the nebula's
[\ion{O}{3}]
emission  originates from the same [\ion{S}{2}] filaments.
From our measured [\ion{S}{2}] $\lambda$6716/$\lambda$6731 intensity ratios, we find a weighted mean value of $1.28\pm0.03$ which indicates mean electron density of 
$180\pm40$ cm$^{-3}$. 

%---------------------------
\section{Discussion}
%---------------------------

Our understanding of the Pa~30's nebula has evolved considerably since its discovery in August 2013 by D.\ Patchick during his WISE search for new Galactic Planetary nebulae.  
Follow-up optical images taken later in 2013 showed no H$\alpha$ emission but a very faint and diffuse [\ion{O}{3} nebula coincident with mid-IR WISE emission
\citep{Kronberger2014}.
Subsequent optical spectra confirmed no detectable H$\alpha$ emission \citep{Gvara2019}
but did later reveal 
[\ion{S}{2}] emission  with radial velocities up to $\pm 1100$ km s$^{-1}$ \citep{Ritter2021}.

No other optical images were reported until 2022 when
\citet{Fesen2023} investigated
Pa~30's optical structure 
through [\ion{S}{2}] emission line images.
Their resulting image revealed an extraordinary and highly structured nebula 
unlike any other Galactic SN remnant, with dozens of radially
aligned filaments and a convergence point near the hot central star.
They proposed the remnant's filaments were the result of wind ablation off ejecta knots due to Pa~30's
unusual central star which central star's 16,000 km s$^{-1}$ wind despite the lack of detected filamentary knots.

%%%**********************************
%%%Figure 4: HST HST Blurred and GMOS
%%%%**********************************
\begin{figure*}[ht]
\begin{center}
\includegraphics[angle=0,width=14.0cm]{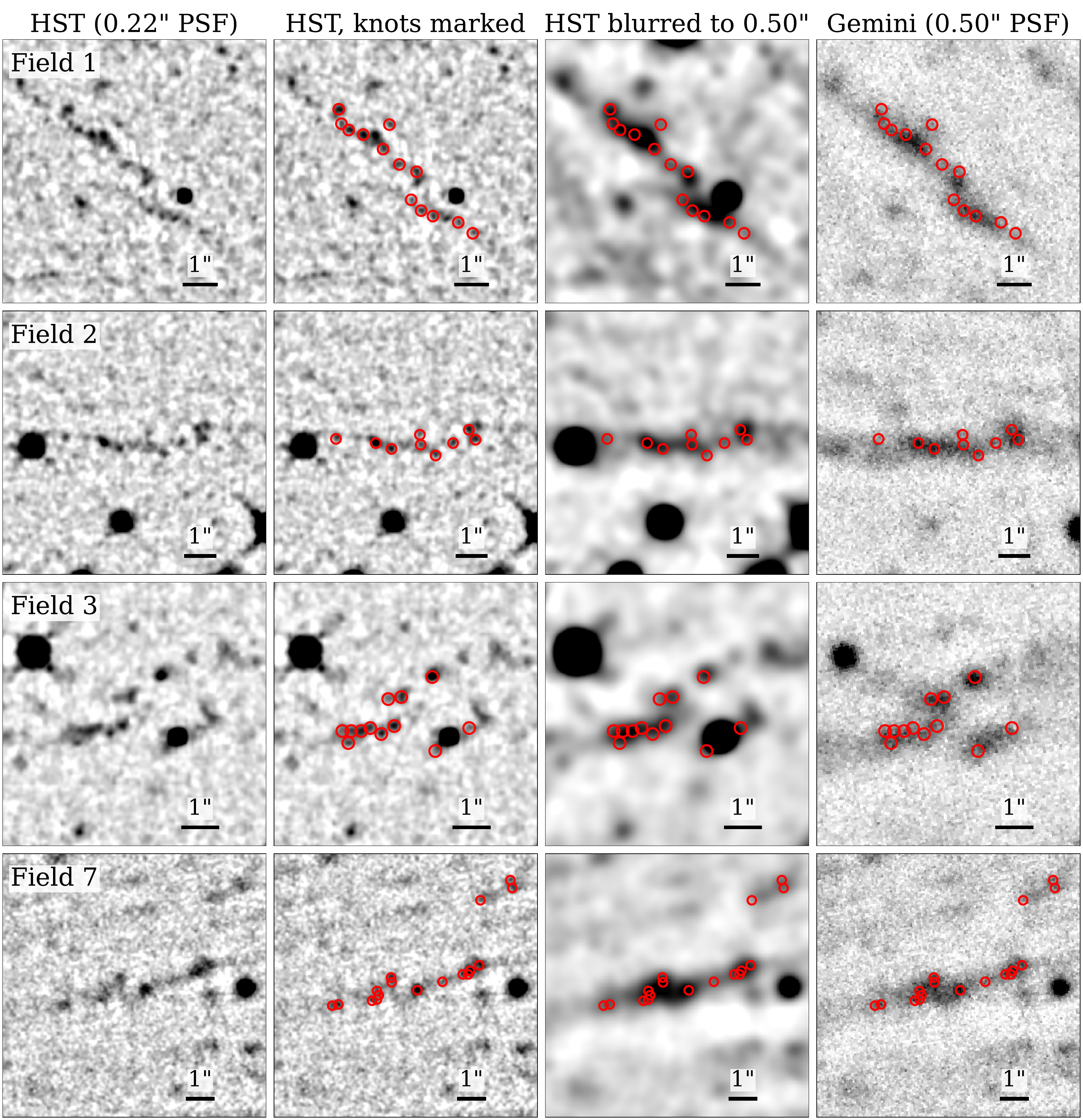} \\
\caption{HST WFC3/IR and Gemini GMOS [\ion{S}{2}] images of four Pa 30 filaments. Columns, left to right: the HST F098M image (PSF = 0.22$''$); the same with the filament knots circled; the HST image blurred to the 0.50$''$ Gemini resolution; and the Gemini GMOS [\ion{S}{2}] image. 
At ground-based resolution, the compact HST knots and the fainter emission around them merge into smooth filaments with apparent $\approx0.6''$ knots like those seen by Gemini.
\label{HST_vs_GMOS}
}
\end{center}
\end{figure*}
%%%%%%%********************************

The recently published 2025 Gemini [\ion{S}{2}] image 
\citep{Cunningham2026} advanced things considerably, detecting nearly an order of magnitude more filaments and concluded that each is made up of streams small knots about 0.6$''$ in size.  They also found that the remnant's faint [\ion{O}{3}] emission filaments are the same ones seen in the [\ion{S}{2}] images.

The progression of increasingly better resolution of Pa~30's filamentary nebula is illustrated in Figure~\ref{res}  where we show MDM, Gemini, and HST images of one of the remnant's brighter filaments. What was seen in the 2022 image as an irregular filament with relatively constant brightness along its length, is seen in the 2025 Gemini image \citep{Cunningham2026} as an irregular chain of knots. These knots turn out to be composed of much smaller knots that are marginally resolved in the 2026 HST image.

Below we show how our HST images reconcile with the knot sizes seen in the recent Gemini image of \citet{Cunningham2026}, and then discuss the implications of our HST results on proposed models of Pa~30 filament formation.

\subsection{Reconciliation of HST and Gemini Images}

Our results do not conflict with those of \citet{Cunningham2026}; rather, the two data sets measure different things. Our HST 
[S~II+III] image shows that filament knots have a median FWHM of $0.154''$, about four times smaller than the $\approx0.6''$ size inferred from the GMOS images.
Blurring HST knots to the Gemini GMOS 0.5$''$ resolution reproduces 0.6$'' - 0.65''$  knot-like features reported by \citet{Cunningham2026}.

This is shown in Figure~\ref{HST_vs_GMOS}. Running from top to bottom, images of four relatively bright Pa~30 filaments are shown here (Fields 1, 2, 3, and 7). In the columns moving from left to right, are the unsmoothed F098M [S II+III] image, then that same image with our identified knots marked with small red circles, next  to that is an image is that same HST image now smoothed to a PSF of 0.5$''$ equal that of \citet{Cunningham2026}'s Gemini GMOS images with the same red knot ID circles. Finally, in the last column, are our identified knot circles overlayed on Gemini images.

\begin{figure}
\begin{center}
\includegraphics[angle=0,width=8.0cm]{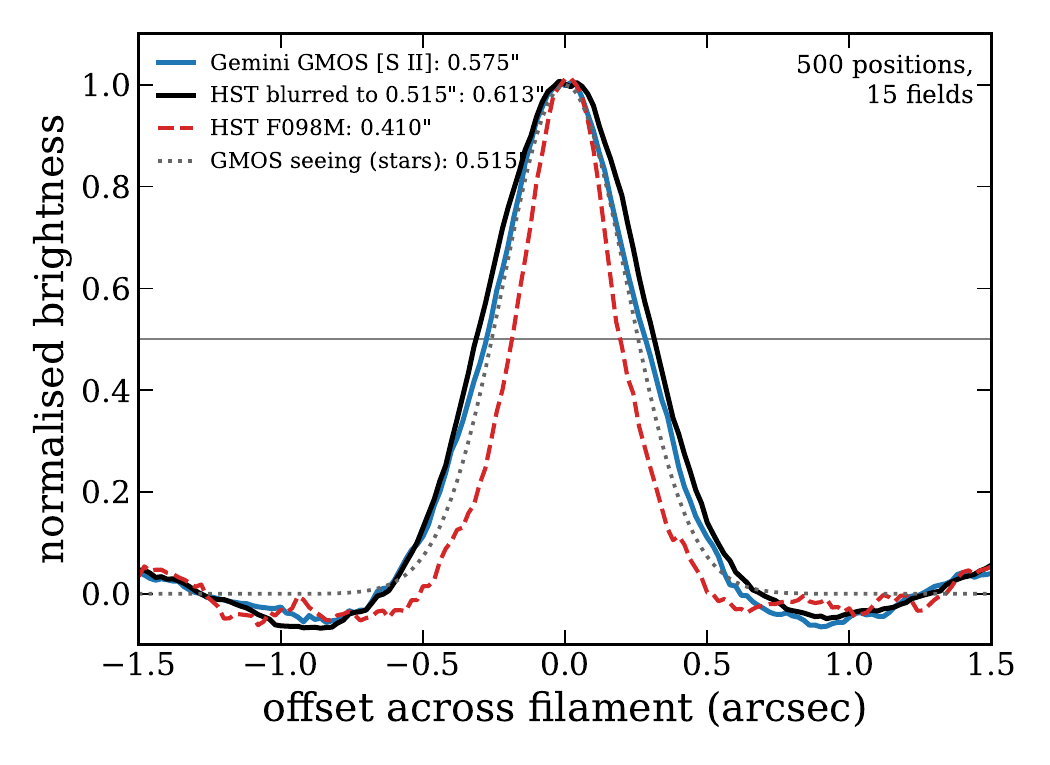} \\
\caption{Cross filament knots of GMOS and blurred HST images
as a complement to the visual comparison shown in Figure 10. 
\label{fig11}
}
\end{center}
\end{figure}

To test this quantitatively, 
we blurred the HST F098M image to the measured GMOS seeing, 
$0.515''$, from 23 compact sources and measured 
cross-filament profiles at 500 positions along the GMOS filament ridges in our 15 fields. Processing the GMOS, blurred-HST and original HST images was done identically
(see Fig.~\ref{fig11}). The median profiles have FWHM of
$0.575'' \pm0.006''$ (GMOS),
$0.613'' \pm 0.009''$ (blurred HST) and $0.410'' \pm 0.010''$ (HST), with a systematic uncertainty of $\sim0.05''$ from the background subtraction. At ground-based resolution, the filament cross-sections are therefore dominated by the seeing; removing it gives an intrinsic filament width of only $\approx0.26''$ (GMOS) to $\approx0.33''$ (blurred HST), consistent with the $\approx0.34''$ obtained directly from the HST image.

Although not all features seen in the Gemini images correspond to our identified knots, the brighter HST knots, when blurred to Gemini resolution, match very well what the Gemini images detected. Some of this difference might be where the HST image is sensitive to knots bright in [\ion{S}{3}] emission.
Similar comparisons through the nebula confirm that our infrared HST image when blurred to Gemini GMOS resolution generate excellent matches.

%%%**********************************
%%%Figure 11: 
%%%%**********************************
%%%%%%%%%%%%%%%%%%%%%%%%%%%%%%%%%%%

Thus our knot analysis and results reconcile the apparent differences in measured Pa~30 knot sizes.
Our infrared HST image better resolves the structure of the nebula's filaments where individual knots are roughly 0.15 arcsec and fainter in surface brightness than the blends suggest, spaced at a minimum of about 
$0.40'' - 0.6''$ within the filaments. Although the Gemini image is much deeper than our WFC3/IR image and thus generates a truly striking image of the Pa 30 nebula, our HST image provides a better look at the remnant's filamentary knot scale.

%---------------------------------------------------------
\subsection{Implications for the Origin of the Filaments}
%---------------------------------------------------------

HST images of young SNRs like Cas A and nova shells such as GK Per have shown that  ejecta knots
with angular sizes below one arcsecond are not uncommon
\citep{Fesen2011, Shara2012}.
What makes the knots in Pa~30 unusual are their remarkable  arrangement in tight streams together with their large population for such a relatively small nebula.

The Gemini image of Pa~30 revealed that the Pa~30 nebula is composed of many hundreds of filaments with typical lengths of 
$5''$ to $15''$. From our HST [S~II+III] image analysis,
we find a median nearest-neighbor separation of $0.6''$ (i.e., $\sim$1.7 knots per arcsec)
along the length of these filaments. Assuming this value and adopting an typical filament length of $\sim10''$, this suggests that  thousands of small knots might be present in the entire nebula, without even taking into account for the likelihood of much fainter knots not counted in our filament knot analysis.

Although several dozens of small ejecta knots have been seen in limited regions of some young SNRs (e.g., the N-rich knots in Cassiopeia A; \citealt{Hammell2008}), nothing like Pa~30's radial filament morphology plus its composition of thousands of compact, sub-arcsecond ejecta knots has been seen before. 

Our finding of such a knot rich nebula may offer clues as to the nebula's formation process. Pa~30's dominate thin filamentary structure naturally leads to suggestions of finger-like
Rayleigh–Taylor (RT) instability features. \citet{Duffel2024} and \citet{Pikus2026} have suggested that highly
efficient cooling in ejecta
will cause a drop of pressure support in Rayleigh-Taylor fingers, 
leading them to be compressed, suppressing the growth of
Kelvin-Helmholtz instability so the RT fingers elongate
into long, thin spikes of material much like that seen in Pa~30.

The elemental abundances in the Pa 30 nebula are poorly known.
\citet{Oskinova2020} found that C, O, Ne and Mg were likely abundant in the nebula based on X-ray data
suggesting a composition relatively rich in
intermediate mass elements (IME).
The fact that only lines of S and Ar are seen in knot optical spectra suggest an unusual composition.
The lack of detected line emissions from 
H$\alpha$, He $\lambda$5876 [\ion{N}{2}] 
$\lambda\lambda$6548,6583,and 
[\ion{Ca}{2}] $\lambda\lambda$7291,7324  suggest a composition
poor in H, He, N, and Ca.

The absence of detected [\ion{O}{1}] $\lambda$6300 and
[\ion{O}{2}] $\lambda\lambda$7319,7330 line emissions
is especially puzzling in the light of the remnant's observed
[\ion{O}{3}] emission.
A high ionization state for the knots that could hide these emission lines seems unlikely given that the low ionization lines of [\ion{S}{2}] are so prominent in their spectrum. 

Also puzzling is the lack of iron and nickel emission lines. For example, in the weakest deflagration SN~Ia models 
\citep{Fink2014}, the mass of $^{56}$Ni and iron group elements is an order of magnitude more abundant
than the mass of IMEs. Yet, our F127M image centered on [\ion{Fe}{2}] 1.2570 $\mu$m, a line commonly seen in both young and old SNRs 
\citep{Oliva1990,Gerardy2001,Kokusho2026}, did not show any appreciable emission.

Likewise, optical spectra of Pa~30 knots do not show 
line emissions from either [\ion{Fe}{2}] 
$\lambda$7155 located right next to the detected 
[\ion{Ar}{3}] $\lambda$7136 line, or [\ion{Ni}{2}] 
$\lambda7376$ like that seen in the
Crab Nebula \citep{MacAlpine1989} that might signal a high Fe or Ni abundance like that predicted by SN Iax models
\citep{Shen2012,Shen2017,Schwab2016,Lach2022}.
Nonetheless, a composition rich in IMEs could help shorten the cooling time scale leading to the formation of streams of small. compact knots within RT formed fingers like that observed.

Following the initial detection of Pa~30 filaments, 
\citet{Fesen2023} proposed  ablation off ejecta knots 
from the WD's  high velocity wind  as the mechanism of forming radially aligned filaments
like that seen in some planetary nebulae. While our HST images 
show some evidence for some faint emission in between individual knots along the lengths of several filaments, the images do not consistently show emission always  out ahead of most filament knots. Any faint emission around the knots appear more consistent with simply neighboring fainter knots rather diffuse ablated material. However, this could be much better explored in deeper images such as with JWST.

A wind-driven origin of the remnant's filaments has been more recently proposed 
by \citet{Coughlin2026}. Their model also invokes RT instabilities for filament creation but here the RT fingers form at the interface of the central star's high speed wind and surrounding CSM. 
The filaments subsequently elongate due to the Kelvin–Helmholtz-stable nature of a large initial density contrast
between the low density of the WD’s wind and CSM.
supplemented by the WD's slowly wind density. 

But it is not clear that this formation models
will generate knotty filaments across the nebula's entire
650 to 1400 km s$^{-1}$ thick shell. They also viewed Pa~30's
filament empty central region (V$_{\rm exp}$ $< 650$ km s$^{-1}$) as more likely a consequence of the finite timescale over which the RT instability operates, rather than a wind
termination shock. On the other hand, it might be that this filament free inner region of the Pa~30 nebula is due to the WD's 16,000 km s$^{-1}$ wind. A similar situation is seen in the Crab Nebula where the Crab pulsar's expanding magnetic field and synchrotron nebula is thought to be the mechanism for the near absence of that remnant's filaments below $\sim$700 km s$^{-1}$ with the notable exception of a thick He band of ejecta centered around the pulsar \citep{Clark1983,MacAlpine1989, Hester2008}.

%%%%%%%%%%%%%%%%%%%%%%%%%%%%%%%%%%%%%%%%%%%%%%%%%%%%
\section{Conclusions}
%%%%%%%%%%%%%%%%%%%%%%%%%%%%%%%%%%%%%%%%%%%%%%%%%%%%

Using near infrared HST WFC3/IR F098M filter images sensitive to [\ion{S}{2}] and 
[\ion{S}{3}] emission lines from the suspected young Galactic supernova remnant, Pa~30, show that its radially align filaments are composed of numerous small knots.  Our findings include:

1) Using an off-band image to discriminate field stars along with recent deep ground-based Gemini [\ion{S}{2}] images to help trace the filaments, a survey of remnant 15 regions was found to contain 292 detected knots, 226 of which were `clean' filament knot detections.

2) The median PSF-corrected FWHM for the clean knots is $0.154''$ ($\sim$350 AU at 2.3 kpc), with 14\% being unresolved
at a PSF of $0.225''$. Tests with over 37,000 artificial knots placed on these filaments indicate that this median is close to the knots' true size, with 81–89\% of the artificial knots recovered in our knot detection method.

3) Pa~30's knots are not round. They are elongated along the lengths of the filaments. Fitting each knot with axes aligned to the local filament direction and calibrating with artificial knots gives intrinsic sizes 
of $\simeq0.22''$ along and by $\simeq0.12''$ across from individual knot fits and $\simeq0.18''$ along and by $\simeq0.11''$ across from knot stacks,
an axis ratio of about 2. 
We find that about 75\% of knots are longer when measured along the filament. 

4) The minimum spacing of knots along a filament 
is typically $0.4'' - 0.6''$ or $\sim 1400$ AU so several times their own size, but can be one arcsecond or larger in some places. This results in discontinuous chains of separate clumps rather than a continuous knotty structure.

5) WFC3/IR images taken using the F127M filter showed no [\ion{Fe}{2}] 1.2 $\mu$m emission from the nebula at a level flux 
above 5\% that of seen using the F098M [S II + III] filter.
This suggests that the remnant's [\ion{Fe}{2}] emissions are quite weak.

6) The density sensitive 
[\ion{S}{2}] 
$\lambda$6716/$\lambda$6731 line ratio 
measured for a small number of knots suggest electron densities $\leq200$ cm$^{-3}$ for the nebula's filamentary knots consistent with previous estimates.

7) Although these results at first seem at odds with the 
conclusion of \citet{Cunningham2026} based on Gemini images that the Pa~30 knots have almost a uniform FWHM = $0.6''$, we find excellent agreement with their data when the HST images are blurred to the $0.5''$ PSF of the Gemini data.
Our measured knots are about 4 times smaller than their quoted $0.6''$ size.
At $0.5''$ seeing, the compact HST knots and the fainter emission around them merge into the $\sim0.6''$ structures seen in the Gemini images.
One caveat is that F098M records  [\ion{S}{2}] 1.03 µm and [\ion{S}{3}] $\lambda$9531, while GMOS records only [\ion{S}{2}] $\lambda\lambda$6716, 6731. The two images could trace slightly different gas with HST seeing only the densest cores. 

The size difference of knots we found across and along is robust but our across-filament width difference between Fields 1–7 and 8–15 has not been checked against distance from the nebula center or position angle. That check would show whether the knots change in width with position in the nebula. Our numbers for knot spacing in filaments should also be treated as a lower limit. For a few  of the remnant's brightest filaments
where knot separations were measured strictly along individual filaments, values were found consistently at the low end of our quoted range, i.e., $\sim 0.4''$. 

Finally, our knot size measurements are just at the limit of our HST WFC3/IR image resolution, so we cannot rule out a large number of even smaller knots when examined at even higher resolution. 
The detection of dozens of bright unresolved knots in our study along with faint unresolved filament emission may signal the presence of a large population of even smaller knots, possibly constituting a power law like knot size distribution.

Two on-going HST and JWST programs targeting Pa~30 may significantly improve our understanding of its nebula.
A Cycle 33 HST program (PI: Cunningham) 
uses WFC3/UVIS with a narrow [\ion{S}{2}] filter, giving about twice the resolution of our WFC3/IR images, and a JWST program (PI: Caiazzo) will obtain complementary  MIRI/IFU observations.
However, a JWST NIRCam imaging program would provide the greatest improvement on determining the remnant's knot emission properties. At 0.9 -- 1.1 $\mu$m, NIRCam's 
diffraction-limited PSF is about 
$0.03'' - 0.04''$.

Such high resolution would make several things possible: 1) Accurate
knot shapes showing perhaps smaller knots with compact heads with tails pointing outward. This would bear directly on ejecta instabilities and wind ablation models, 2) With a 6 times larger collecting area, JWST's NIRCam would likely detected 
both smaller knots and fainter filament emission in between knots, thereby
improving completeness and knot spacing statistics, and 3) 
Given the remnant's expansion $\sim 1000$ km s$^{-1}$, outer knots will shift roughly 0.09$''$ per year allowing for direct measurement of any 
remnant acceleration due the central star's high velocity wind.
Because the Pa~30 nebula is the closest remnant of what is believed a rare Type Iax supernova, further studies of it are clearly worthwhile. \\

We wish to thank Mike Shull and Brad Schaeffer
for helpful comments and suggestions,
and Alison Vick at STScI for helping with the observing plan and schedule.
We also thank Tim Cunningham for sending us his Gemini imaging paper on Pa~30 in advance of its publication as it was helpful in our image analysis.
This work was supported by NASA through the Guest Observer grant
GO-18127 from the Space Telescope Science Institute (STScI), which is operated by the Association of Universities for Research in Astronomy, Inc.
We used Claude AI Opus 5.5 (Anthropic) to develop and run the knot detection, classification, artificial-knot, stacking and photometry analyses, and generated graphs and tables that improved the communication of our analysis.
The authors are entirely responsible for the accuracy of all figures, graphs, text, and references in the final manuscript. 

\facilities{MDM (OSMOS), HST (WFC3/IR), LBT (MODS)}

\software{PYRAF \citep{Green2012}, SAOImage (DS9; \citep{Joye2003}), XMGrace, 
numpy \citep{vanderWalt2011}, scipy \citep{Virtanen2020}, 
pandas \citep{McKinney2010}, matplotlib \citep{Hunter2007}, scikit-image \citep{vanderWalt2014} } \\

\clearpage

\begin{appendix}
%\subsection{LBT [\ion{S}{2}] Emission Image and Electron Density Sensitive Ratio.}

\begin{figure}[ht!]
\begin{center}
\includegraphics[angle=0,width=15.0cm]{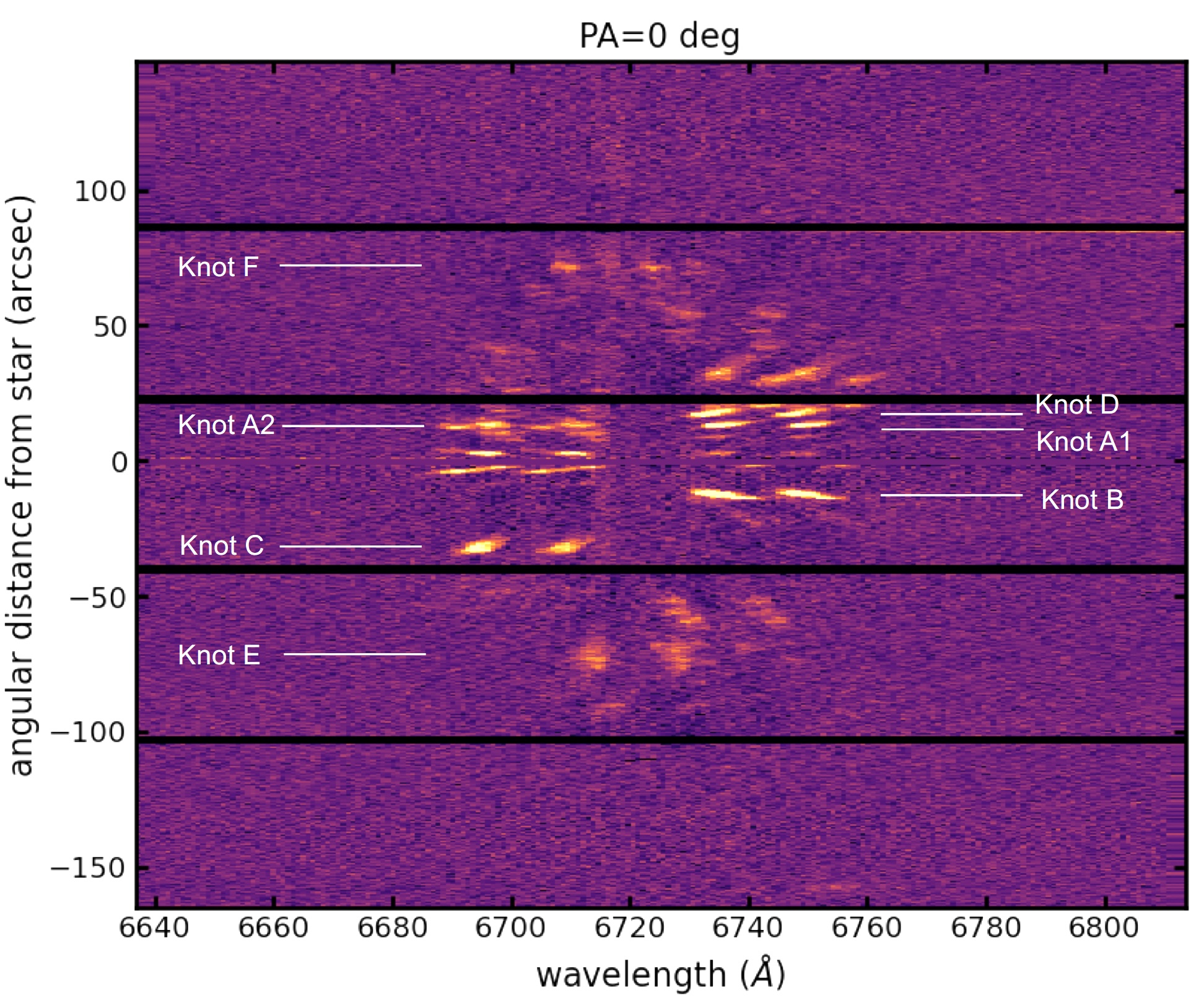} \\
\caption{{\bf {Top:}} 2D composite image of Pa~30's [\ion{S}{2}] $\lambda\lambda$6716,6731 emission lines from LBT/MODS spectra taken  at PA=0. Remnant knots where the 6716/6731 intensity ratios were measured (Table 4) are indicated. Telluric emission features and stellar continua have been subtracted. The black horizontal bands are slit mask supports.
\label{figA1}
}
\end{center}
\end{figure}

As described in Section~3.5, the long-slit LBT spectra intersected knots over the entire radius of Pa30. Figure~\ref{figA1} displays a portion of the two-dimensional spectrum at [\ion{S}{2}] with the knots measured in Table~\ref{spec} marked. The spectrum displays the geometry of the expanding remnant with velocities ranging from 600~\kms\ to more than 1200~\kms . That there are no low-velocity emissions detected near the stellar remnant indicates a central hole in the remnant's  [\ion{S}{2}] emission with a radius of $650$~\kms\ \citep{Fesen2023,Cunningham2024} . At large angular separations, the expansion is perpendicular to our line-of-sight, so the radial velocity approaches zero. 

Individual knots are extended in both the spatial and dispersion directions and appear to point to the stellar remnant. This results from the homologously expanding strings of knots where their velocities increase with distance from the central star. 

In general, the $\lambda 6716$ component appears brighter than $\lambda 6731$ emission. This immediately implies an electron density that is less than $10^3$~cm$^{-3}$. The two lines are sufficiently close that correcting for the large
dust reddening increases the 6716/6731 ratio by only 0.6\% . 

We use the {\it pyNeb} code \citep{luridiana2015} to calculate the [\ion{S}{2}] intensity ratio as a function of electron density and temperature. The electron density for each knot is given in Table~\ref{spec}\ for $T_e=10000^\circ$K. Taking a weighted mean of the observed line ratios gives an electron density of 180$\pm 40$~cm$^{-3}$.

%%%**********************************
%%%Figure A1: 
%%%%**********************************
\end{appendix}

\bibliography{aas_ref2.bib}
\bibliographystyle{aasjournal}

\end{document}